\documentclass[aps,prb,twocolumn,amsmath,floatfix]{revtex4-1}
\usepackage{dcolumn}    
\usepackage{bm}         
\usepackage{hyperref}   
\usepackage[usenames,dvipsnames]{xcolor}
\usepackage[normalem]{ulem} 
\usepackage{graphicx}
\usepackage[english]{babel}

\usepackage[utf8]{inputenc}
\usepackage[T1]{fontenc}

\usepackage{scalerel,stackengine}
\stackMath
\newcommand\reallywidehat[1]{%
\savestack{\tmpbox}{\stretchto{%
  \scaleto{%
    \scalerel*[\widthof{\ensuremath{#1}}]{\kern-.6pt\bigwedge\kern-.6pt}%
    {\rule[-\textheight/2]{1ex}{\textheight}}
  }{\textheight}%
}{0.5ex}}%
\stackon[1pt]{#1}{\tmpbox}%
}

\newcommand{\angstrom}{\text{\normalfont\AA}}
\newcommand{\specialcell}[2][c]{\begin{tabular}[#1]{@{}c@{}}#2\end{tabular}}

\newcommand*{\x}{\bm{x}}

\newcommand*{\Vnl}{V_{\rm I}}

\begin{document}

\title{Kolmogorov and Kelvin wave cascades in a generalized model for quantum turbulence}

\author{Nicol\'as P. M\"uller}
\author{Giorgio Krstulovic}
\affiliation{%
  Universit\'e C\^ote d'Azur, Observatoire de la C\^ote d'Azur, CNRS,
  Laboratoire Lagrange,  Boulevard de l'Observatoire CS 34229 - F 06304 NICE Cedex 4, France
}


\begin{abstract}
        We performed numerical simulations of decaying quantum turbulence by
        using a generalized Gross-Pitaevskii equation, that includes a beyond
        mean field correction and a nonlocal interaction potential. The
        nonlocal potential is chosen in order to mimic He II by introducing a
        roton minimum in the excitation spectrum. We observe that at large
        scales the statistical behavior of the flow is independent of the
        interaction potential, but at scales smaller than the intervortex
        distance a Kelvin wave cascade is enhanced in the generalized model. In
        this range, the incompressible kinetic energy spectrum obeys the weak
        wave turbulence prediction for Kelvin wave cascade not only for the
        scaling with wave numbers but also for the energy flux and the
        intervortex distance. 
\end{abstract}

\maketitle

\section{Introduction}

One of the most fundamental phase transitions in low temperature physics is the Bose-Einstein condensation \cite{pitaevskii_boseeinstein_2016}. It occurs when a fluid composed of bosons is cooled down below a critical temperature. In that state, the system has long-range order and can be described by a macroscopic wave function. One of the most remarkable properties of a Bose-Einstein condensate (BEC) is that it flows with no viscosity. Well before the first experimental realization of a BEC by Anderson et al. \cite{anderson_observation_1995}, Kaptiza and Allen discovered that helium becomes superfluid below 2.17K \cite{kapitza_viscosity_1938, allen_flow_1938}. A couple of years later, London suggested that superfluidity is intimately linked to the phenomenon of Bose-Einstein condensation~\cite{london_lphenomenon_1938}. Since then, superfluid helium and BECs made of atomic gases have been extensively studied, both theoretical and experimentally. In particular, the fluid dynamics aspect of quantum fluids has renewed interest due the impressive experimental progress of the last fifteen years. Today it is possible to visualize and follow the dynamics of quantum vortices, one the most fundamental excitations of a quantum fluid \cite{fonda_direct_2014,serafini_dynamics_2015}.

Quantum vortices are topological defects of the macroscopic wave function describing the superfluid. They are nodal lines of the wave function and they manifest points and filaments in two and three dimensions respectively. To ensure the monodromy of the wave function, vortices have the topological constraint that the circulation (contour integral) of the flow around the vortex must be a multiple of the Feynman-Onsager quantum of circulation $\kappa=h/m$, where $h$ is the Planck constant and $m$ is the mass of the Bosons constituting the fluid \cite{pitaevskii_boseeinstein_2016}. In superfluid helium their core size is of the order of 1\AA\,whereas in atomic BECs is typically of the order of microns \cite{barenghi_introduction_2014}. Quantum vortices interact with other vortices similarly to the classical ones. They move thanks to their self-induced velocity and interact with each other by hydrodynamics laws \cite{schwarz_threedimensional_1988}. Unlike ideal classical vortices described by Euler equations, quantum vortices can reconnect and change their topology despite the lack of viscosity of the fluid in which they are immersed \cite{koplik_vortex_1993}.

At scales much larger than the mean intervortex distance $\ell$, the quantum nature of vortices is not very important as many individual vortices contribute to the flow. One could expect then that the flow is similar, in some sense, to classical one. Indeed, if energy is injected at large scales, a classical Kolmogorov turbulent regime emerges. Such a regime has been observed numerically \cite{nore_decaying_1997, baggaley_vortexdensity_2012, shukla_quantitative_2019} and experimentally in superfluid helium \cite{maurer_local_1998, salort_energy_2012}.  In a three-dimensional turbulent flow, energy is transferred towards small scales in a cascade process \cite{frisch_turbulence_1995}. 
In a low temperature turbulent superfluid, when energy reaches the intervortex distance, energy keeps being transferred to even smaller scales where it can be efficiently dissipated by sound emission. The mechanisms responsible for this are the vortex reconnections and the wave turbulence cascade of Kelvin waves, that have its origin in the quantum nature of vortices \cite{vinen_decay_2001}.

Describing a turbulent superfluid is not an easy task, in particular for superfluid helium. One of the main reasons is the gigantic scale separation existing between the vortex core size and the typical size of experiments, currently of the order centimeters or even meters \cite{rousset_superfluid_2014}. Their theoretical description began at the beginning of the 20th century by the pioneering works of Landau and Tisza where superfluid helium was modeled by two immiscible fluid components \cite{donnelly_quantized_1991}. In this two-fluid model, the thermal excitations constitute the so called normal fluid that is modeled through the Navier-Stokes equations whereas a superfluid component is treated as an inviscid fluid. It was later realized that the thermal excitations interact with superfluid vortices through a scattering process that leads to a coupling of both component by mutual friction forces \cite{donnelly_quantized_1991}. Today the two-fluid description, known as the Hall-Vinen-Bekarevich-Khalatnikov model is understood as a coarse-grained model where scales smaller than the intervortex distance are not considered. The quantum nature aspects of superfluid vortices are therefore lost. However, this model remains useful for describing the large scale dynamics of finite temperature superfluid helium. An alternative model was introduced by Schwarz \cite{schwarz_threedimensional_1988}, where vortices are described by vortex filaments interacting through regularized Biot-Savart integrals. However, the reconnection process between lines needs to be modeled in an ad-hoc manner and by construction the model excludes the dynamics of a superfluid at scales smaller than the vortex core size. Finally, in the limit of low temperature and weakly interacting BECs, a model of different nature can be formally derived which is the Gross-Pitaevskii (GP) equation, obtained from a mean field theory \cite{pitaevskii_boseeinstein_2016}. This model naturally contains vortex reconnections \cite{koplik_vortex_1993,villois_universal_2017}, sound emission \cite{krstulovic_radiation_2008,villois_irreversible_2020} and is known to also exhibit a Kolmogorov turbulent regime at scales much larger than the intervortex distance \cite{nore_decaying_1997}. Although this model is expected to provide some qualitative description of superfluid helium at low temperatures, it lacks of several physical ingredients. For instance, in GP, density excitations do not present any roton minimum as it does in superfluid helium, where interaction between boson are known to be much stronger than in GP \cite{donnelly_observed_1998}. However, there have been some successful attempts to include such effect in the GP model. For instance, a roton minimum can be easily introduced in GP by using a nonlocal potential that models a long-range interaction between bosons \cite{pomeau_model_1993,berloff_motions_1999, reneuve_structure_2018}. The stronger interaction of helium can also be included phenomenologically by introducing high-order terms in the GP Hamiltonian. Note that these terms can be derived as beyond mean field corrections \cite{lee_eigenvalues_1957}. Some generalized version of the GP model has been used to study the vortex solutions \cite{berloff_modeling_2014,villerot_static_2012} and some dynamical aspects such as vortex reconnections \cite{reneuve_structure_2018}. Naively, for a turbulent superfluid, we can expect that such generalization of the GP model might be important at scales smaller than the intervortex distance and with less influence at scales at which Kolmogorov turbulence is observed.

In this work, we study quantum turbulent flows by performing numerical simulations of a generalized Gross-Pitaevskii (gGP) equation. We compare the effect of high-order nonlinear terms and the effect of a nonlocal interaction potential in the development and decay of turbulence at scales both larger and smaller than the intervortex distance. Remarkably, by modeling superfluid helium with a nonlocal interaction potential and including high-order terms, the range where a Kelvin wave cascade is observed is extended and becomes manifest. Using the dissipation (or rate of transfer) of incompressible kinetic energy we are able to show that the weak wave turbulence results \cite{lvov_spectrum_2010} are valid not only to predict the scaling with wave number but also with the energy flux and the intervortex distance.

The manuscript is organized as follows. Section \ref{sec:teo} introduces the gGP model and discusses its basic properties and solutions. It also discusses how the vortex profile is modified in this generalized model. All useful definitions to study turbulence are also given here. Section \ref{sec:simulations} gives a brief overview of the predictions of quantum turbulence and the numerical methods used in this work. Also, it includes the results of different simulations at moderate and high resolutions by varying the different parameters of the beyond mean field correction and the introduction of a nonlocal potential. Finally in Section \ref{sec:conclusions} we present our conclusions.


\section{Theoretical description of superfluid turbulence}
\label{sec:teo}

On this section we introduce the generalized Gross-Pitaevskii model used in this work. We also discuss and review some of the basic properties of the model as its elementary excitations and its hydrodynamic description.

\subsection{Model} \label{subsec:model}

The Gross-Pitaevskii equation describes the low temperature dynamics of weakly interacting
bosons of mass $m$
\begin{equation}
i\hbar\frac{\partial\psi}{\partial t} = -\frac{\hbar^2}{2m}\nabla^2\psi - \mu
\psi + g|\psi|^2\psi,
\label{eq:GP}
\end{equation}
 where $\psi$ is the condensate wave function, $\mu$ the chemical
potential, $\hbar=h/(2\pi)$ and $g=4\pi \hbar^2 a_s/m$ is the coupling constant fixed by the
$s$-wave scattering length $a_s$ that models a local interaction between
bosons . Note that the use of a local
potential assumes a weak interaction between bosons, which certainly is not the
case for other systems like He II and for dipolar gases
\cite{lahaye_physics_2009}.

A generalized model that is able to describe more complex systems can be
obtained by considering a nonlocal interaction between bosons. With proper modeling \cite{pomeau_model_1993,berloff_motions_1999,reneuve_structure_2018}, density excitations exhibit a roton minimum in their spectrum as the one observed in He II \cite{donnelly_observed_1998}.
It also describes well the behavior of dipolar condensates \cite{griesmaier_generation_2007, santos_rotonmaxon_2003}.
In helium and other superfluids, the interaction between bosons is stronger and high-order nonlinearities are needed for proper modeling. For instance, in helium high-order terms are considered to mimic its equations of state \cite{berloff_motions_1999} and in dipolar BECs beyond mean field terms are needed to describe the physics of recent supersolid experiments \cite{roccuzzo_rotating_2020}.

We consider the generalized Gross-Pitaevskii (gGP) model written as
\begin{eqnarray}
\label{eq:gGPE_hbar} i\hbar\frac{\partial\psi}{\partial t} &= &-\frac{\hbar^2}{2m}\nabla^2\psi
    - \mu (1 + \chi)\psi \\
\nonumber    &&+ g\left( \int \Vnl(\bm{x}-\bm{y})|\psi(\bm{y})|^2 \mathrm{d}^3y\right)\psi + g\chi
    \frac{|\psi|^{2(1+\gamma)}}{n_0^{\gamma}} \psi.
\end{eqnarray}
%
%
where $\gamma$ and $\chi$ are two dimensionless parameters that
determine the order and amplitude of the high-order terms, respectively. The interaction potential $V_{\rm I}$ is normalized such that $\int \Vnl(\x)\mathrm{d}^3x=1$. The chemical potential and the interaction coefficient of the high-order term have been renormalized such that $|\psi_0|^2=n_0=\mu/g$ is the density of particles for the ground state of the system for all values of parameters. The GP equation \eqref{eq:GP} is recovered by simply setting $\Vnl(\bm{x} - \bm{y}) = \delta(\bm{x} - \bm{y})$ and $\chi=0$.

The gGP equation is not intended to be a first principle model of superfluid helium, but it has the advantage of at least introducing in a phenomenological manner some important physical aspects of helium.

\subsection{Density waves}
\label{subsec:density_waves}

The dispersion relation of the GP model is easily obtained by linearizing equation \eqref{eq:GP} about the ground state. The Bogoliubov dispersion reads
\begin{equation}
\omega_B(k) = c_0k\sqrt{\frac{\xi_0^2k^2}{2} + 1},
\label{eq:Bog}
\end{equation}

\noindent where $k$ is the wave number, $c_0=\sqrt{gn_0/m}$ is the speed of
sound of the superfluid and $\xi_0=\hbar/\sqrt{2mgn_0}$ is the healing
length at which dispersive effects become important. The healing length also fixes the vortex core size.

A similar calculation leads to the Bogoliubov dispersion relation in the case of the gGP model \eqref{eq:gGPE_hbar}
\begin{equation}
    \omega(k) = ck \sqrt{\frac{\xi^2k^2}{2} + \frac{\hat{\Vnl}(k) +
    \chi(\gamma+1)}{1+\chi(\gamma+1)}},
\label{eq:dispersion}
\end{equation}
where $\hat{\Vnl}(\bm{k}) = \int e^{i\bm{k}\cdot\bm{r}}\Vnl(\bm{r})\mathrm{d}^3r$
is the Fourier transform of the interaction potential normalized such that
$\hat{\Vnl}(k=0)=1$. The inclusion of beyond mean-field terms and a nonlocal potential yields to a renormalized speed of sound and healing length. They are given
in terms of $c_0$ and $\xi_0$ by
\begin{align}
    c   &= c_0\sqrt{1+\chi(\gamma+1)} \label{eq:c}\\
    \xi &= \frac{\xi_0}{\sqrt{1+\chi(\gamma+1)}}.
\label{eq:healing}
\end{align}
Note that, in what concerns low amplitude density waves, the effect of high-order terms is a simple renormalization of the healing length and the speed of sound. Depending on the shape and properties of the nonlocal potential,  the dynamics and steady solutions can be drastically modified. Note that the product between $c$ and $\xi$ remains constant
because it is related to the quantum of circulation $\kappa= h/m = c\xi 2\pi\sqrt{2}=c_0\xi_0 2\pi\sqrt{2}$.

In order to be able to compare the systems with different type of interactions, it is convenient to rewrite Eq. \eqref{eq:gGPE_hbar} in terms of its intrinsic length $\xi$ and speed of sound $c$ and the bulk density $n_0$. The gGP model then becomes
\begin{widetext}
\begin{equation}
    \partial_t\psi = -i\frac{c}{\xi\sqrt{2}(1+\chi(\gamma+1))}\bigg[-(1+\chi(\gamma+1))\xi^2\nabla^2\psi - (1 +
    \chi) \psi + \chi \frac{|\psi|^{2(1+\gamma)}}{n_0^{1+\gamma}}\psi +
    \frac{\psi}{n_0} \int \Vnl(\bm{x}-\bm{y}) |\psi(\bm{y})|^2 \mathrm{d}^3y \bigg].
\label{eq:gGPE}
\end{equation}
\end{widetext}
In numerical simulations we will express lengths in unit of the healing length $\xi$. A natural time scale to study excitations is the fast turnover time $\tau=\xi/c$. However, this small-scale based time is not appropriate for turbulent flows. For such flows, it is customary to use the large-eddy turnover time corresponding to the typical time of the largest coherent vortex structure and will be defined later.

\subsubsection{Modeling superfluid helium excitations}

In this work, we aim at mimicking some properties of superfluid helium II, in particular its roton minimum  in the dispersion relation. 
For the sake of simplicity, we use an isotropic nonlocal interaction potential used in previous works
\cite{berloff_modeling_2014,reneuve_structure_2018}. With our normalization it reads

\begin{equation}
    \hat{\Vnl}(\bm{k}) = \left[1 - V_1 \left(\frac{k}{k_{\mathrm{rot}}}\right)^2 +
    V_2 \left(\frac{k}{k_{\mathrm{rot}}}\right)^4\right]\exp
    \left(-\frac{k^2}{2k_{\mathrm{rot}}^2}\right),
\label{eq:potential}
\end{equation}
where $k_{\mathrm{rot}}$ is the wave number associated with the roton
minimum and $V_1\geq0$ and $V_2\geq0$ are dimensionless parameters to be
adjusted to mimic the experimental dispersion relation of helium II \cite{donnelly_observed_1998}.
The effects of different functional forms of the nonlocal potential have been
studied in previous works, showing that only a phase-shift of $\psi$ and the
overall amplitude of the density depend on the precise form of the interaction
\cite{villerot_static_2012}.

\begin{figure}[tpb]
        \includegraphics[width=.99\linewidth, trim={1cm 1.5cm 1cm 2cm}, clip]{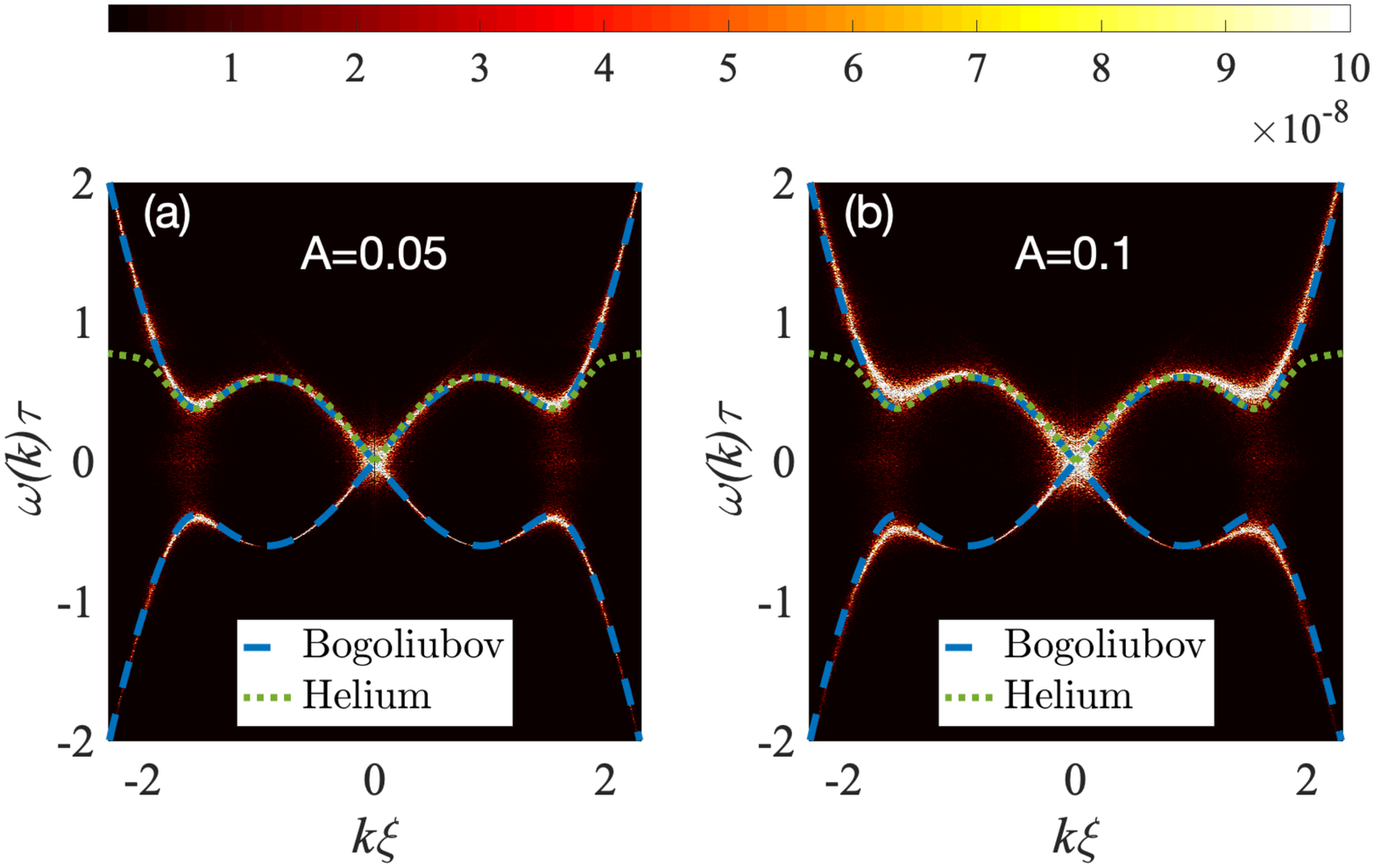}
  \caption{Spatiotemporal dispersion relation for simulations with $1024^2$
          grid points with a nonlocal potential and beyond mean field corrections. Light zones correspond to excited frequencies. Figures (a) and (b) correspond to different amplitude of the perturbation $A$, both exhibiting a roton minimum. Experimental observations (green dotted
          line, see \cite{donnelly_observed_1998}) and theoretical dispersion
          following equation \eqref{eq:dispersion} (blue dashed line) are
          shown.}
  \label{fig:dispersion}
\end{figure}

In order to compare the dispersion relation \eqref{eq:dispersion} with the experimental data \cite{donnelly_observed_1998}, we plot the helium dispersion relation in units of the helium healing length $\xi_{\rm He}=0.8\angstrom$ and its turnover time $\tau_{\rm He}=\xi_{\rm He}/c_{\rm He}=3.36\times10^{-13}$s, where $c_{\rm He}=238$ m/s is the speed of sound in He II. The measured helium dispersion relation is displayed in Fig. \ref{fig:dispersion} as green dotted lines.

It was reported in Reneuve et al. \cite{reneuve_structure_2018} that introducing a roton minimum in the GP dispersion relation (without beyond mean field terms) that matches helium measurements leads to an unphysical crystallization under dynamical evolution of a vortex. We confirm such behavior in our simulations. In order to avoid such spurious effect of the model, in reference \cite{reneuve_structure_2018} the frequency associated to the roton minimum was set to higher values to be able to study vortex reconnections. 
We have numerically observed that this crystallization takes place even
when a first order correction of the beyond mean field expansion is
included, with values of $\chi=0.1$ and $\gamma=1$. For this reason, we chose
a higher order expansion with $\gamma=2.8$ for the simulations with a nonlocal
potential, value that was already used in the literature to study the vortex
density profile in superfluid helium \cite{berloff_motions_1999}.
Furthermore, with this value no crystallization is observed for all test
cases and all the simulations performed in this work.

The dispersion relation of a nonlinear wave system can be measured numerically by computing the spatiotemporal spectrum of the wave field \cite{shukla_quantitative_2019}. As an example, in Fig. \ref{fig:dispersion} we also display the spatiotemporal spectrum of small density perturbations of a numerical simulation of the gGP model with $1024^2$ collocation points and with parameters set to $\gamma=2.8$, $\chi=0.1$, $V_1=4.54$, $V_2=0.01$ and
$k_{\mathrm{rot}}\xi=1.638$ (see details on numerics later in Sec. \ref{sec:simulations}), for two different amplitude values $A$. Dark zones indicate that no frequencies are excited, while light zones correspond to the excited ones with the total sum normalized to one. The parameters have been set in a way such that they match qualitatively the dispersion relation measured in helium. As expected for weak amplitude waves, the numerical and theoretical dispersion relations coincide. For larger wave amplitudes, theoretical prediction \eqref{eq:dispersion} and numerical measurements slightly differ together with an apparent broadening of the curve. This is a typical behavior of nonlinear wave systems \cite{nazarenko_wave_2011}. In the following sections, all simulations with a nonlocal interaction are performed with aforementioned set of parameters.

\subsection{Hydrodynamic description}
\label{subsec:hydro}

The GP equation maps into an hydrodynamic description by introducing the Madelung
transformation

\begin{equation}
\psi = \sqrt{\rho/m} \exp\left(\frac{i\phi}{\sqrt{2}c\xi} \right),
\label{eq:madelung}
\end{equation}

\noindent which allows the mapping of the wave function with the fluid mass
density $\rho = m|\psi|^2$ and with the fluid velocity $\bm{v} = \bm{\nabla}
\phi$. Replacing equation \eqref{eq:madelung} into the gGP model \eqref{eq:gGPE} two
hydrodynamic equations are obtained

\begin{align}
&\frac{\partial \rho}{\partial t} + \bm{\nabla}\cdot(\rho\bm{v}) = 0 \label{eq:continuity} \\
 &   \frac{\partial \phi}{\partial t} + \frac{1}{2}(\bm{\nabla}\phi)^2 =-h[\rho] +(c\xi)^2\frac{\nabla^2\sqrt{\rho}}{\sqrt{\rho}}, \label{eq:euler}
\end{align}
with
\begin{align}
 & h[\rho] = -c_0^2(1+\chi) +c_0^2\frac{\Vnl*\rho}{\rho_0} + c_0^2 \chi\left(\frac{\rho}{\rho_0} \right)^{\gamma+1}.
\end{align}

\noindent Here $*$ denotes is the convolution product and $\rho_0 = m|\psi_0|^2$ is the fluid mass density of the ground state. These correspond to the continuity and Bernoulli equations, respectively, of a fluid with an enthalpy per unit of mass $h[\rho]$ \cite{nore_decaying_1997}. The last term of equation \eqref{eq:euler} is called
the quantum pressure. Note that hydrodynamic pressure is given by
\begin{equation}
p[\rho]=\frac{c_0^2\rho}{\rho_0}\left[ \frac{1}{2}\Vnl*\rho +\chi\frac{\gamma+1}{\gamma+2}\frac{\rho^{\gamma+1}}{\rho_0^\gamma}\right].
\end{equation}
As expected, for large amplitude waves, the speed of sound reads $\left.\frac{\partial p}{\partial \rho}\right|_{\rho_0}=c_0^2(1+\chi(\gamma+1))=c^2$.

Although the fluid is potential, it admits vortices as topological defects of
the wave function. A stationary vortex solution of \eqref{eq:gGPE} is a zero of
the wave function where the circulation around it is quantized with values $\pm
s\kappa$ with $s$ an integer. Because of this last condition, topological
defects are also called quantum vortices.

A quantum vortex has vortex core size of order $\xi$ and depends on the parameters of the gGP model. By replacing
the Madelung transformation \eqref{eq:madelung} into the gGP equation
\eqref{eq:gGPE} and solving in cylindrical coordinates, a differential equation
for the vortex profile is directly obtained
\begin{equation}
\begin{aligned}
        \frac{1}{r}\frac{\mathrm{d}}{\mathrm{d}r} \left(r\frac{\mathrm{d}R}{\mathrm{d}r}\right)+ \bigg\{ &1-\frac{s^2\xi_0^2}{r^2}- \Vnl*R^2 + \\
                                                          &+ \chi(1- R^{2\gamma+2}) \bigg\}\frac{R}{\xi_0^2} = 0
\end{aligned}
\end{equation}

\noindent where $R(r) = \sqrt{\rho(r)/\rho_0}$ defines the density profile of the
vortex line in the radial direction $r$.

\begin{figure}
  \centering
  \includegraphics[width=1\linewidth]{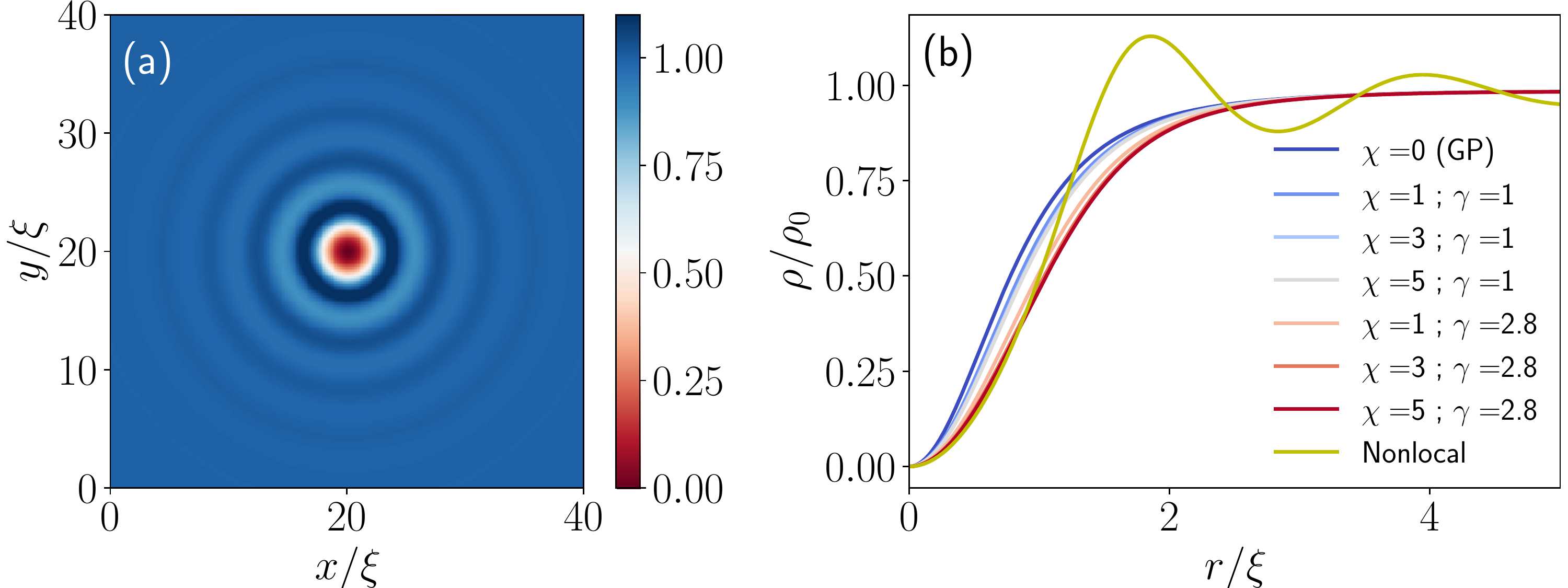}
  \caption{(a) Mass density of a two-dimensional vortex with a nonlocal
  potential. (b) Density profile of a vortex for the gGP model with different values of
  the nonlinearity and a local potential, and a single profile with a nonlocal potential (yellow line). The vortex core
  size tends to increase with the nonlinearity.}
  \label{fig:profile}
\end{figure}

Figure \ref{fig:profile} (a) displays the mass density of a two-dimensional vortex in the case where the
nonlocal interaction potential is included.
The roton minimum introduces some density fluctuations around the center of the vortex which is a well-known
pattern. Such a behavior has already been studied before,
for example its interaction with an obstacle \cite{pomeau_model_1993}, the
dynamics of vortex rings \cite{berloff_motions_1999} and in reconnection
processes \cite{reneuve_structure_2018}. Figure \ref{fig:profile} (b) shows the radial dependence of the density profile of a vortex for different parameters of the gGP model. Numerical simulations were performed with $4096^2$ grid
points with standard numerical methods (see section \ref{subsec:numerical} for
details). Even though all curves tend to collapse when plotted as a function of the healing length $\xi$, the vortex core size slightly increases (in units of $\xi$) when
the nonlinearity of the system is increased. Note that for the present range of parameters, $\xi_0/\xi$ varies in the range $(1,4.4)$. The relatively good collapse of the vortex core size thus justifies the choice of $\xi$ to parametrize the gGP model while varying the beyond mean field parameters.

\subsection{Energy decomposition and helicity in superfluids}
\label{subsec:decomposition}

It is convenient to write the free energy per unit of mass $\mathcal{F}$ of a quantum fluid such that it vanishes when evaluated in the ground state of the system ($\psi=\sqrt{\rho_0/m}=\sqrt{n_0}$). For the gGP model in equation \eqref{eq:gGPE}, it is given by
\begin{eqnarray}
        \nonumber    \mathcal{F} = \frac{c_0^2}{n_0V} \int \Bigg[&& \xi_0^2 |\nabla\psi|^2 +
    \frac{|\psi|^2}{2n_0}(\Vnl*|\psi|^2) - (1 + \chi)|\psi|^2 + \\
    && \frac{\chi |\psi|^{2(\gamma+2)}}{n_0^{\gamma+1}(\gamma+2)} + 
    \frac{n_0}{2} + \chi n_0 \frac{\gamma+1}{\gamma+2} \Bigg]\mathrm{d}^3{r},
\label{eq:E}
\end{eqnarray}

\noindent with $V$ the volume of the fluid. Following standard procedures applied in simulations of GP quantum turbulence \cite{nore_decaying_1997}, the free energy can be decomposed as  $\mathcal{F}=E_{\rm kin}^I+E_{\rm kin}^C+E_{\rm q}+E_{\rm int}$ where $E_{\mathrm{kin}}^{\rm I} = \frac{1}{V\rho_0} \int {\left([\sqrt{\rho}\bm{v}]^{\rm I}\right)} ^2\mathrm{d}^3{r}$, $E_{\mathrm{kin}}^{\rm C} =  \frac{1}{V\rho_0} \int {\left([\sqrt{\rho}\bm{v}]^{\rm C}\right)} ^2\mathrm{d}^3{r}$ and $E_{\rm q}= \frac{c^2\xi^2}{V\rho_0}\int |\bm{\nabla} \sqrt{\rho}|^2 \mathrm{d}^3{r}$, with $[\sqrt{\rho}\bm{v}]^{\rm I}$ the regularized incompressible velocity obtained via the Helmholtz decomposition and $[\sqrt{\rho}\bm{v}]^{\rm C}=\sqrt{\rho}\bm{v}-[\sqrt{\rho}\bm{v}]^{\rm I}$ the compressible one. The internal energy per unit of volume is defined in the gGP model as
\begin{align}
        \nonumber    E_{\mathrm{int}} = &\frac{c_0^2}{V\rho_0} \int \bigg[ \frac{1}{2\rho_0} (\rho-\rho_0)\Vnl*(\rho-\rho_0) \\
&+  \left(\frac{\rho}{\rho_0}\right)^{\gamma+1}\frac{\chi\rho}{\gamma+2} - \chi \rho +\chi \frac{\gamma+1}{\gamma+2}\rho_0\bigg] \mathrm{d}^3{r}.
\label{eq:Eint}
\end{align}
Note that $E_{\mathrm{int}}=0$ if $\rho=\rho_0$.
The corresponding energy spectra are defined in a straightforward way for the quadratic quantities \cite{nore_decaying_1997}. For the internal energy spectrum, it is defined as follows
\begin{align}
\nonumber    E_{\mathrm{int}}(k) = &\frac{c_0^2}{V\rho_0} \int \Bigg[
        \frac{1}{2\rho_0}\reallywidehat{(\rho-\rho_0)}_{-{\bf k}}\hat{\Vnl}(k)
    \reallywidehat{(\rho-\rho_0)}_{{\bf k}} \\
    &+ \frac{\chi}{\gamma+2}
    \hat{\rho}_{-{\bf k}}\reallywidehat{ \left(\frac{\rho}{\rho_0}\right)_{{\bf k}}^{\gamma+1}} +
    \reallywidehat{(\chi\rho_0\frac{\gamma+1}{\gamma+2} - \chi \rho)}_{{\bf k}}\Bigg]
    \mathrm{d}\Omega_k ,
\label{eq:Eint_spec}
\end{align}
where $\mathrm{d}\Omega_k$ is the element of surface of the shell $|{\bf k}|=k$
where the hat stands for the Fourier transform defined in the same way as in
the nonlocal potential after equation \eqref{eq:dispersion}. Note that this
particular choice of the spectrum is not unique and has been made so that the
ground state $\rho=\rho_0$ contributes with no internal energy to the system.
It is also worth noting that with this definition, the internal energy spectrum
may take negative values.

Besides the energies, there is another quantity in quantum turbulence that
presents a great interest in the dynamics of quantum vortices \cite{clarkdileoni_helicity_2016, zuccher_helicity_2015, scheeler_helicity_2014}, which is the central line helicity per unit of volume
\begin{equation}
        H_{\rm c} = \frac{1}{V}\int \bm{v}(\bm{r}) \cdot \bm{\omega}(\bm{r}) \mathrm{d}^3{r}.
        \label{eq:Helicity}
\end{equation}
Note that $V H_{\rm c}/\kappa^2$ is the total number of helicity quanta.
Formally, this quantity is ill defined for a quantum vortex as the vorticity is $\delta$-supported on the filaments and the velocity is not defined on the vortex core. However, in the GP formalism, this singularity can be removed by taking proper limits \cite{clarkdileoni_helicity_2016}. We use the definition central line helicity proposed in reference \cite{clarkdileoni_helicity_2016} as its numerical implementation is tedious but straightforward and well behaved for vortex tangles.

\section{Evolution of Quantum turbulent flows}
\label{sec:simulations}

This section gives a brief overview about the predictions in quantum turbulence both at large and small scales and details of the numerical methods used to run the simulations. There is also a description of the flow visualization in the presence of a nonlocal interaction potential, and the results of the flow evolution at moderate and high resolution are shown. In particular, it is studied the dependence of the different components of the energy and the helicity with beyond mean field parameters and with the introduction of a nonlocal interaction potential.

\subsection{A brief overview of cascades in quantum turbulence\label{subsec:cascades}}

Quantum turbulence is characterized by the disordered and chaotic motion of a
superfluid. Energy injected, or initially contained, at large scales is transferred towards small scales in a Richardson cascade process \cite{frisch_turbulence_1995}.
In the context of GP turbulence, the contribution of vortices to the global energy can be studied by looking at the incompressible kinetic energy $E_{\mathrm{kin}}^I$ and its associated spectrum. As the system evolves, vortices interact transferring energy between scales. Besides, the incompressible kinetic energy is transferred to the quantum, internal and compressible energy through vortex reconnections and sound emission \cite{vinen_decay_2001,villois_irreversible_2020}. After some time, acoustic excitations thermalize and act as a thermal bath providing a (pseudo) dissipative mechanism, so vortices shrink until they vanish \cite{krstulovic_anomalous_2011,krstulovic_energy_2011,barenghi_introduction_2014}.

Three-dimensional quantum turbulence presents two main statistical properties.
At scales much larger than the intervortex distance $\ell$, but much smaller than the integral scale $L_0$, the quantum character of vortices is not important and we can think as the system being coarse-grained. At such scales the system presents a behavior that resembles to classical turbulence with a direct energy cascade, that is the transfer of energy from large to small structures. As a consequence, in this range, the
incompressible kinetic energy spectrum $E_{\mathrm{kin}}^I(k)$ follows the Kolmogorov
prediction \cite{frisch_turbulence_1995,nore_decaying_1997, nemirovskii_quantum_2013, skrbek_developed_2012}
\begin{equation}
        E_{\mathrm{kin}}^I(k) = C_K \epsilon^{2/3}k^{-5/3},
        \label{eq:K41}
\end{equation}
\noindent where $C_K\sim1$ and $\epsilon$ is the dissipation rate of
the flow, which in GP quantum turbulence is associated with the rate of change of incompressible kinetic energy $\epsilon=-\mathrm{d}E_{\mathrm{kin}}^I/\mathrm{d}t$ that is expressed in units of $[\epsilon] = Length^2/Time^3$.

In classical three-dimensional inviscid flows, helicity (\ref{eq:Helicity}) is also conserved. Associated to this invariant, a second direct cascade is expected to be also present at large scales, obeying the scaling \cite{brissaud_helicity_1973}
\begin{equation}
        H(k) = C_H \eta \epsilon^{-1/3} k^{-5/3}, 
        \label{eq:hel_spec}
\end{equation}
\noindent where $C_H\sim1$ and $\eta=-\mathrm{d}H/\mathrm{d}t$ is the dissipation rate of helicity. This dual cascade has been also observed in quantum turbulent flows described by the GP equation \cite{clarkdileoni_dual_2017}.

At scales smaller than the intervortex distance, each quantum vortex can be thought as if it were
isolated. Hence, its behavior can be described by the wave turbulence theory as such vortices admit hydrodynamic excitations known as Kelvin waves. Such waves propagate along vortices and interact nonlinearly among themselves. As a result, energy is transferred towards small scales through a process that can be described by the theory of weak wave turbulence \cite{nazarenko_wave_2011}. An agitated debate arose concerning the prediction of the energy spectrum. Two independent groups leaded by L'vov \& Nazarenko \cite{lvov_spectrum_2010} and Kozik \& Svistunov \cite{kozik_theory_2009}, starting from the same equations and applying the same theory derived different predictions. Even though, today there is more numerical data supporting L'vov \& Nazarenko prediction \cite{krstulovic_kelvinwave_2012,boue_exact_2011,baggaley_kelvinwave_2014,villois_evolution_2016}, this issue is still debated \cite{eltsov_amplitude_2020,sonin_comment_2020,eltsov_reply_2020}. We present here the L'vov\&Nazarenko prediction as, we will see later, it was found to be in agreement with our numerical data. This theoretical prediction is derived for an almost straight vortex of period $L_{\rm v}$ and, as discussed in reference \cite{eltsov_amplitude_2020}, some care is needed in order to apply the model to a turbulent vortex tangle. We partially reproduce here and adapt to our case the considerations of reference \cite{eltsov_amplitude_2020}. The wave turbulence L'vov \& Nazarenko prediction is
\begin{equation}
        e_{\mathrm{KW}}(k) = C_{\mathrm{LN}}
        \frac{\kappa\Lambda\epsilon_{\rm KW}^{1/3}}{\Psi^{2/3}k^{5/3}},
        \label{eq:KW_badunits}
\end{equation}
with $\Lambda = \mathrm{log}(\ell/\xi)$ and $C_{\mathrm{LN}}\approx 0.304$ \cite{boue_exact_2011}. Here $\epsilon_{\rm KW}=-\mathrm{d}e_{\mathrm{KW}}/\mathrm{d}t$ is the mean energy flux
per unit of length $L_{\rm v}$ and density $\rho_0$. Note their respective dimensions are $[\epsilon_{\rm KW}]=Length^4/Time^3$ and $[e_{\mathrm{KW}}(k)]=Length^5/Time^2$. The dimensionless number $\Psi$ is given by
\begin{equation}
        \Psi = \frac{(12\pi C_{\mathrm{LN}})^{3/5}
        {\epsilon_{\rm KW}}^{1/5}}{\kappa^{3/5}k_{\mathrm{min}}^{2/5}} = 
        C_{\mathrm{LN}}^{3/5} \tilde{\Psi},
        \label{eq:Psi}
\end{equation}
 where $k_{\mathrm{min}}$ is the smallest wave number of the Kelvin waves, that
 can be associated with the wave number of the intervortex distance
 $k_{\ell}=2\pi/\ell$ in the case of a vortex tangle
 \cite{eltsov_amplitude_2020}.  $\tilde{\Psi}$ is defined so that it is
 independent of the constant $C_{\mathrm{LN}}$ and proportional to $\Psi$.

In order to compare this result with the incompressible kinetic energy, one can notice that the total energy of Kelvin waves is $L_{\rm v}\rho_0\int e_{\mathrm{KW}}(k)\mathrm{d}k$, where now $L_{\rm v}$ is taken as the total vortex length in the system. As in a turbulent tangle the total vortex length is related to the mean intervortex distance by $L_{\rm v}=V\ell^{-2}$, it follows that the mean kinetic energy spectrum per unit of mass is given by $E_{\mathrm{KW}}(k)=e_{\mathrm{KW}}(k) \ell^{-2}$. The same logic relates the energy flux $\epsilon_{\rm KW}$ of the Kelvin wave cascade to the global energy flux $\epsilon$ of a tangle by  $\epsilon_{\rm KW} = \epsilon \ell^2$ \cite{eltsov_amplitude_2020}. It follows from \eqref{eq:KW_badunits} and the previous considerations that
\begin{equation}
        E_{\mathrm{KW}}(k) = C_{\mathrm{LN}}^{3/5}
        \frac{\kappa\Lambda\epsilon^{1/3}\ell^{-4/3}}{\tilde{\Psi}^{2/3}k^{5/3}}.
        \label{eq:KW}
\end{equation}
Here we have made the assumption that the energy flux in the Kolmogorov range is the same as in the Kelvin wave cascade. This strong assumption might be questioned as energy could be already dissipated into sound by vortex reconnections at different scales diminishing this value \cite{villois_irreversible_2020,proment_matching_2020}. Such extra sinks of energy are difficult to quantify and we will not take them into account. Finally, note that the theory of wave turbulence also predicts the value of the constant $C_{\rm LN}$ \cite{boue_exact_2011}, however in \eqref{eq:KW} several phenomenological considerations have been made and we do not expect an exact agreement. Nevertheless, the scaling with the global energy flux should remain valid.

\subsection{Numerical methods}
\label{subsec:numerical}

We perform numerical simulations of equation \eqref{eq:gGPE} using a
pseudo-spectral method for the spatial resolution applying the  ``$2/3$ rule''
for dealiasing \cite{gottlieb1977numerical}, and a Runge-Kutta method of fourth
order for the time stepping.  The nonlinear term is dealiased twice following
the scheme presented in \cite{krstulovic_energy_2011} in order to also conserve
momentum. Note that in the case of a nonlocal potential, this extra step has no
extra numerical cost.  All simulations were performed in a cubic $L$-periodic
domain.

To observe a Kolmogorov range in GP turbulence it is customary to start from an initial vortex configuration with a minimal acoustic contribution.  The initial condition for the wave function is obtained by a minimization process such that the resulting flow is as close as possible to the targeted velocity field \cite{nore_decaying_1997}. In this work we study the quantum Arnold-Bertrami-Childress (ABC) flow \cite{clarkdileoni_dual_2017}. It is obtained from the velocity field ${\bm{v}}_{\mathrm{ABC}} = \bm{v}_{\mathrm{ABC}}^{(k_1)} +
\bm{v}_{\mathrm{ABC}}^{(k_2)}$, where each ABC flow is given by
\begin{eqnarray}
&\bm{v}_{\mathrm{ABC}}^{(k)} = [B \cos(ky) + C \sin(kz)]\hat{x} + \\
\nonumber &[C \cos(kz) + A \sin(kx)]\hat{y} + [A \cos(kx) + B \sin(ky)]\hat{z}.
\label{eq:ABC}
\end{eqnarray}
We set in this work $(A,B,C) = V_{\rm amp}\,(0.9,1,1.1)/\sqrt{3}$, with $V_{\rm amp}=0.5\,c$. Each ABC flow is a $L$-periodic
stationary solution of the Euler equation with maximal helicity, in the sense that $\bm{\nabla}\times \bm{v}_{\mathrm{ABC}}^{(k)}=k\bm{v}_{\mathrm{ABC}}^{(k)}$. The mean kinetic energy associated with ${\bm{v}}_{\mathrm{ABC}} $ is $E_{\rm kin}^{\mathrm{ABC}}=(A^2+B^2+C^2)=0.2517 c^2$.
Following reference \cite{clarkdileoni_dual_2017}, the wave
function associated to this ABC flow is generated as $\psi_{\mathrm{ABC}} =
\psi_{\mathrm{ABC}}^{(k_1)} \times \psi_{\mathrm{ABC}}^{(k_2)}$, where each mode is
constructed as the product $\psi_{\mathrm{ABC}}^{(k)} = \psi_{A,k}^{x,y,z}
\times \psi_{B,k}^{y,z,x} \times \psi_{C,k}^{z,x,y}$ with
\begin{equation}
        \psi_{A,k}^{x,y,z} = \exp \left\{i \left[\frac{A \sin(kx)}{c\xi\sqrt{2}}\right] \frac{2\pi y}{L}
        + i \left[\frac{A \cos(kx)}{c\xi\sqrt{2}}\right] \frac{ 2\pi z}{L} \right\}
\label{eq:psiABC}
\end{equation}
where the brackets $\left[\,\,\right]$ indicate the integer closest to
the value to ensure periodicity.  This ansatz gives a good approximation for the phase of the initial condition. In order to set properly the mass density
and the vortex profiles, it is necessary to first evolve
$\psi_{\mathrm{ABC}}$ using the generalized Advected Real Ginzburg-Landau equation (imaginary time evolution in a locally Galilean transformed system of reference) \cite{nore_decaying_1997}
\begin{eqnarray}
\nonumber    \partial_t\psi = & -\frac{c_0}{\xi_0\sqrt{2}}\bigg\{-\xi_0^2\nabla^2\psi - (1 +
    \chi) \psi + \chi \frac{|\psi|^{2(1+\gamma)}}{\rho_0^{1+\gamma}}\psi + \\
    & \frac{\psi}{\rho_0}(V * |\psi|^2)  \bigg\} - i \bm{v}_{\mathrm{ABC}}
    \cdot \bm{\nabla} \psi - \frac{(\bm{v}_{\mathrm{ABC}})^2}{2\sqrt{2}c\xi}\psi.
\label{eq:hoARGLE}
\end{eqnarray}
This equation is dissipative and its final state contains a minimal amount of compressible modes. This state is used as initial condition for the gGP equation. Unless stated otherwise, we use a flow at the largest scales of the systems by setting $k_1=2\pi/L$ and $k_2=4\pi/L$ throughout this work.

The numerical simulations performed in this work are summarized in Table \ref{tab:simulations} and regrouped in two different sets.
The first set of simulations (runs A1 - A8) have been performed at a moderate spatial resolution of $N^3=256^3$ grid points to study the effects introduced by the beyond mean field interactions and a nonlocal potential. Each of them has a different value of $\chi$ and $\gamma$ with a local
potential and were compared with a single simulation with a nonlocal interaction potential.
The second set (runs B1 - B6) has been performed to study the scaling of the energy spectra. In these runs, we used a spatial resolution of $512^3$ and
$1024^3$ grid points, different scale separations and initial conditions. These results were also compared with the GP model.
\begin{table}
\centering
\begin{tabular}{c@{\hspace{1pt}}||c|c|c|c|c|c}
     & $N$  &$\chi$     &$\gamma$ 	& $L/\xi$ 	& $\tilde{k}_1,\tilde{k}_2$ 	& \specialcell{Interaction\\potential} \\ \hline \hline
A1 & 256  & 0  		& 1       		& 171     	&  1,2  	  	&  local \\
A2 & 256  & 1    	& 1       		& 171     	&  1,2  		&  local \\
A3 & 256  & 3    	& 1      		& 171     	&  1,2  		&  local \\
A4 & 256  & 5    	& 1       		& 171     	&  1,2  		&  local \\
A5 & 256  & 1    	& 2.8     		& 171     	&  1,2  		&  local \\
A6 & 256  & 3    	& 2.8     		& 171     	&  1,2  		&  local \\
A7 & 256  & 5    	& 2.8     		& 171     	&  1,2  		&  local \\
A8 & 256  & 0.1  	& 2.8     		& 171     	&  1,2  		&  nonlocal \\ \hline
B1 & 512  & 0    	& 1      		& 341     	&  1,2  		&  local \\
B2 & 512  & 0.1  	& 2.8     		& 171     	&  1,2  		&  nonlocal \\
B3 & 512  & 0.1  	& 2.8     		& 341     	&  1,2  		&  nonlocal \\
B4 & 512  & 0.1  	& 2.8     		& 341     	&  2,3  		&  nonlocal \\
B5 & 512  & 0.1  	& 2.8     		& 341     	&  3,4  		&  nonlocal \\
B6 & 1024 & 0.1  	& 2.8     		& 683     	&  1,2  		&  nonlocal \\
\end{tabular}
\caption{Table with the parameters of the different simulations.
$N$ is the linear spatial resolution, $\chi$ and $\gamma$ are the amplitude and order
of the beyond mean field interactions, $L/\xi$ is the scale separation between
the domain size $L$ and the healing length $\xi$, $\tilde{k}_1=k_1 L/2\pi$ and $\tilde{k}_2=k_2 L/2\pi$ are the two wave
numbers where the energy is concentrated for the initial condition, and a local
or a nonlocal interaction potential is used in each of them.}
\label{tab:simulations}
\end{table}

\subsection{Flow visualization}
\label{subsec:visualization}

\begin{figure*}[tpb]
        \centering
        \includegraphics[width=.98\linewidth, trim={0cm 1cm 0cm 1cm}, clip]{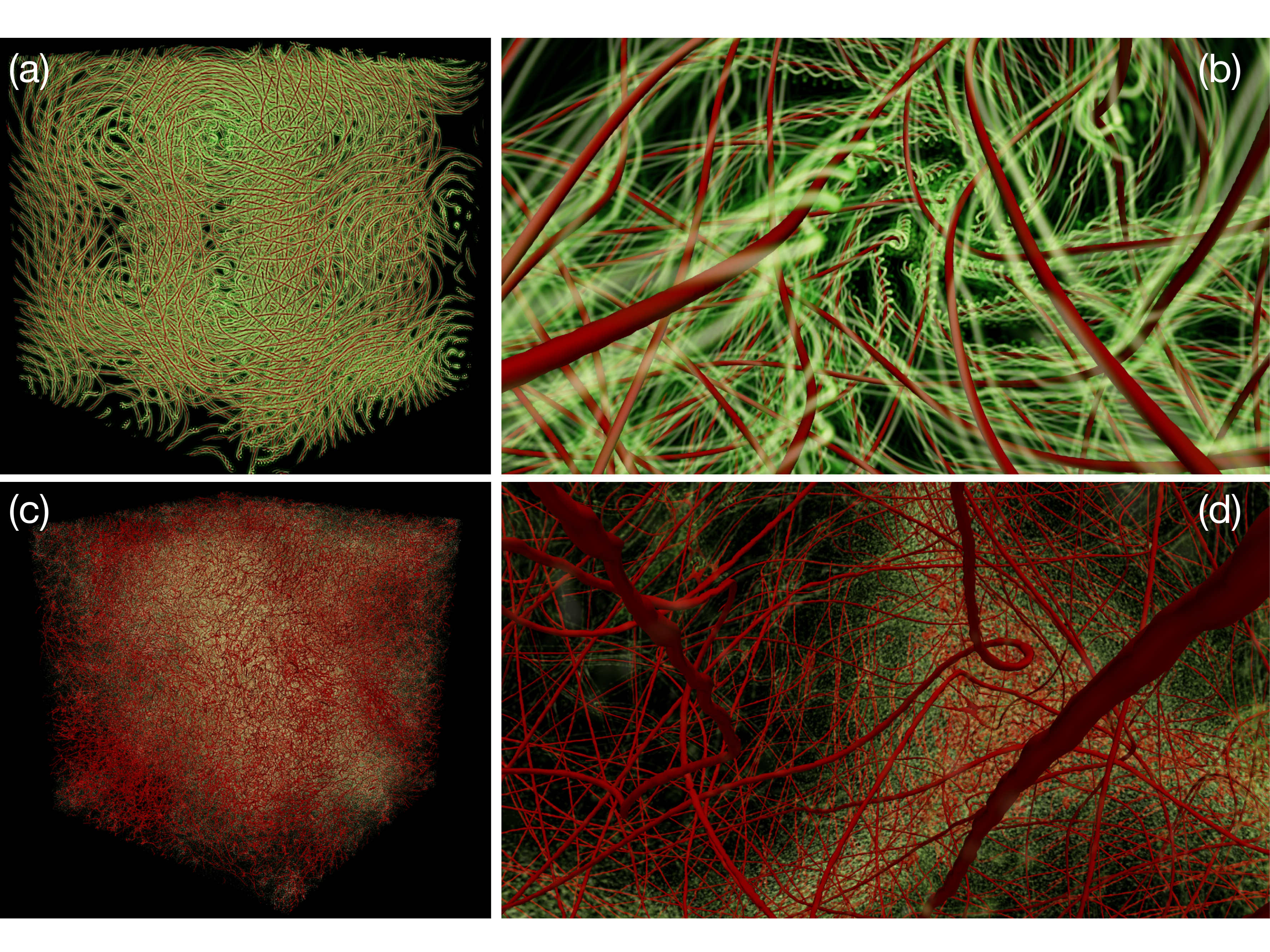}
\caption{(a)-(b) Visualization of an ABC flow at $t=0$ and (c)-(d) for
        $t=1.25\tau_L$ and $t=0.4\tau_L$ respectively, for a resolution of
        $1024^3$ grid points with a nonlocal potential. The isosurfaces of a
        small value the mass density shown in red correspond to the vortex
        lines, and in green are the values of the density fluctuations above
        $\rho_0$. }%
        \label{fig:visu}
\end{figure*}

The introduction of a nonlocal potential, as mentioned in Sec. \ref{subsec:hydro}, allows the
system to reproduce the roton minimum in the excitation spectrum (see Fig. \ref{fig:dispersion}). As a consequence, the density profiles close to the quantum vortices have some fluctuations around the bulk value $\rho_0$ (see Fig. \ref{fig:profile}). These oscillations have been studied for the profile of a two-dimensional vortex \cite{berloff_motions_1999,villerot_static_2012} and have been also observed in three dimensions during vortex reconnections \cite{reneuve_structure_2018}. In the case of a helical vortex tangle, the roton minimum induces a remarkable pattern of density fluctuations around a vortex line. A visualization of the initial condition $\psi_{\rm ABC}$ for run B6 is displayed in Fig. \ref{fig:visu} (a)-(b).
The red structures are isosurfaces of low density values $\rho=0.1\rho_0$ and thus represent the vortex lines. The greenish rendering displays density fluctuations of the field above the bulk value $\rho_0$, that are only observed in the case of a nonlocal potential. In Fig. \ref{fig:visu} (a) we recognize the large scale structures of the ABC flow accompanied by some density fluctuations around the nodal lines. Figure \ref{fig:visu} (b) displays a zoom of the tangle where such fluctuations	are clearly observed. Unlike the (local) GP model, density variations around a vortex line have a very specific pattern, rolling around the nodal lines in a helicoidal manner. Such pattern is a consequence of the maximal helicity initial condition produced by the ABC flow. Indeed, we have also produced a Taylor-Green initial condition \cite{nore_decaying_1997}, that has no mean helicity, and such helicoidal patterns in the density fluctuations are absent, although they are nevertheless developed after some vortex reconnections, as observed in reference \cite{reneuve_structure_2018} (data not shown).
Finally, in Fig. \ref{fig:visu} (c)-(d) we display visualizations of the field
for times $t=1.25\tau_L$ and $t=0.4\tau_L$ respectively. 
Time is expressed in units of the
large-eddy turnover time $\tau_L = L_0/v_{\mathrm{rms}}$ with $v_{\mathrm{rms}}
= \sqrt{2E_{\mathrm{kin}}^I(t=0)/3}$ and $L_0$ its integral length scale given
by $L_0=2\pi/k_2$ with $k_2$ the largest wave number used to generate the initial condition. 
$t=1.25\tau_L$
corresponds to a time when turbulence is developed
, and
$t=0.4\tau_L$ to an early stage of the turbulent development for a
better insight of the flow. As the system evolves, acoustic emissions are
produced and the density fluctuations increase. In Fig. \ref{fig:visu} (c) we
observe a turbulent tangle where a large scale structure is predominant. Figure
\ref{fig:visu} (d) displays a zoom where reconnections and Kelvin waves
propagating along vortices are clearly visible.

\subsection{Temporal evolution of global quantities}
\label{subsec:ResultsI}

In this section we study the behavior of the global quantities of an
ABC flow described by gGP model \eqref{eq:gGPE} with both local and nonlocal
potentials corresponding to runs A in Table \ref{tab:simulations}.

\begin{figure}[tpb]
  \centering
  \includegraphics[width=1\linewidth]{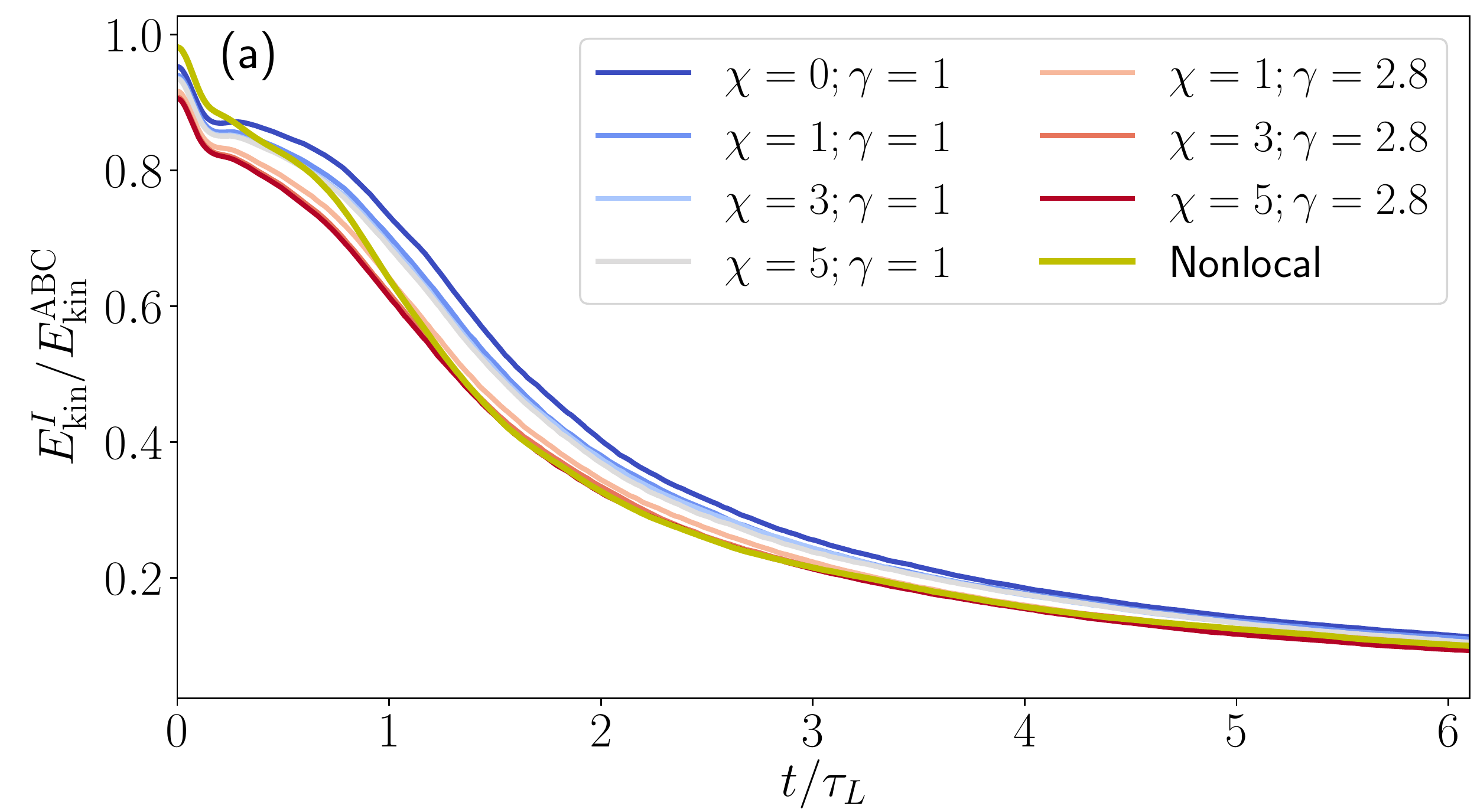}
  \includegraphics[width=1\linewidth]{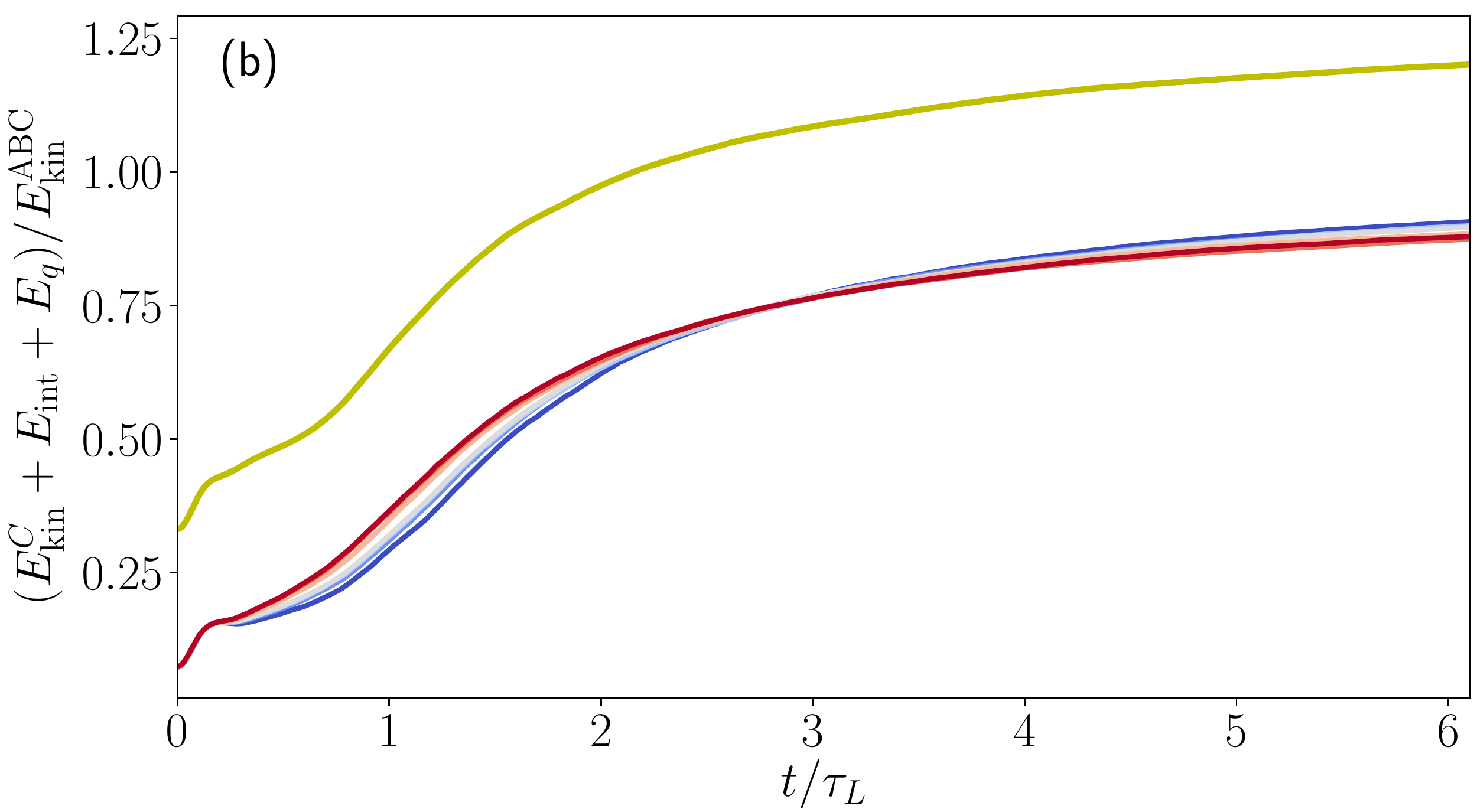}
  \includegraphics[width=1\linewidth]{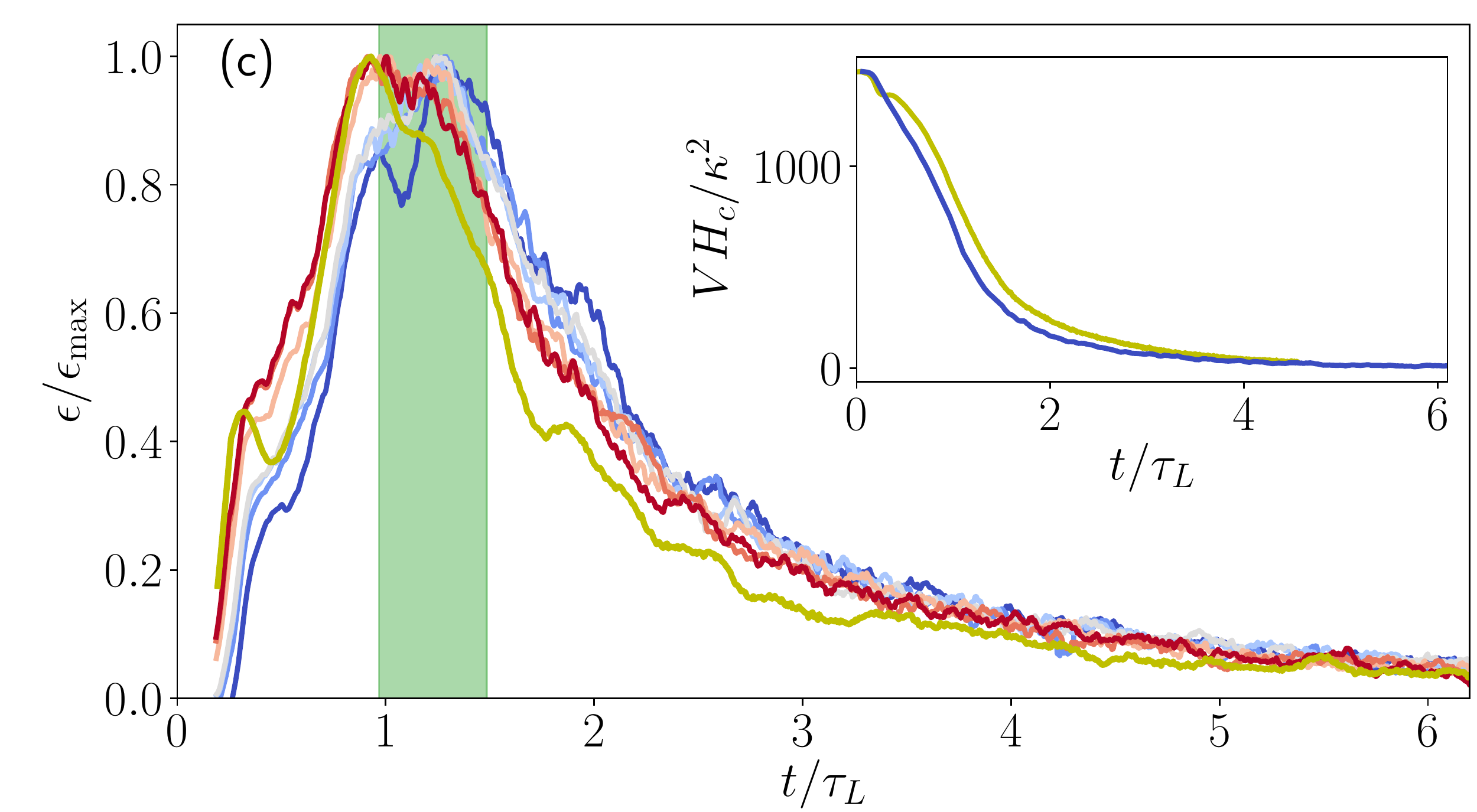}
  \caption{Time evolution of the (a) incompressible kinetic energy, (b) the sum
  of the internal, quantum and compressible kinetic energy and (c) the
  dissipation rate of incompressible energy for runs A in Table
  \ref{tab:simulations}. The inset in (c) shows the evolution of the central
  line helicity. The green area corresponds to the window where the dissipation achieves a maximum and where the time averages were performed.}
  \label{fig:energy}
\end{figure}

Figure \ref{fig:energy} shows the time evolution of the (a) incompressible
kinetic energy and (b) the sum of the quantum, internal and compressible
kinetic components to the total energy. 
We notice that in Fig. \ref{fig:energy} (a) the values of
amplitude and exponent of the beyond mean field interaction and the inclusion
of roton minimum (Runs A1-A8) have a negligible impact on the incompressible
energy of the initial condition, and their effect is very small during the
temporal evolution.
On the other hand, as the fluid can be considered to be more incompressible due to stronger interactions, the density variations respect to the bulk value $\rho_0$ yield larger values of the other energy component between initial times and $t\approx 3\tau_L$ as displayed in Fig. \ref{fig:energy} (b). In particular, for the case of a nonlocal potential the larger values develop through the whole run. Nevertheless, for all runs during the first large-eddy turnover times the main contribution to energy comes from vortices. At later times, energy from vortices is converted into sound. As stated in Sec. \ref{subsec:cascades}, the decay of the incompressible energy can be used to estimate the energy dissipation rate $\epsilon$. Its temporal evolution is displayed in Fig. \ref{fig:energy} (c). As in classical decaying turbulent flows, for quantum flows the Kolmogorov regime is more developed at times slightly after the maximum of dissipation is reached. The green zone in the figure depicts the temporal window where the system is considered to be in a quasi-steady state and a temporal average can be performed to improve statistics. The inset of Fig. \ref{fig:energy} (c) displays that the decay of the central line helicity is independent of the parameters of the gGP model and is consistent with the one reported in reference \cite{clarkdileoni_dual_2017}. 

\begin{figure}[tpb]
  \centering
  \includegraphics[width=1\linewidth]{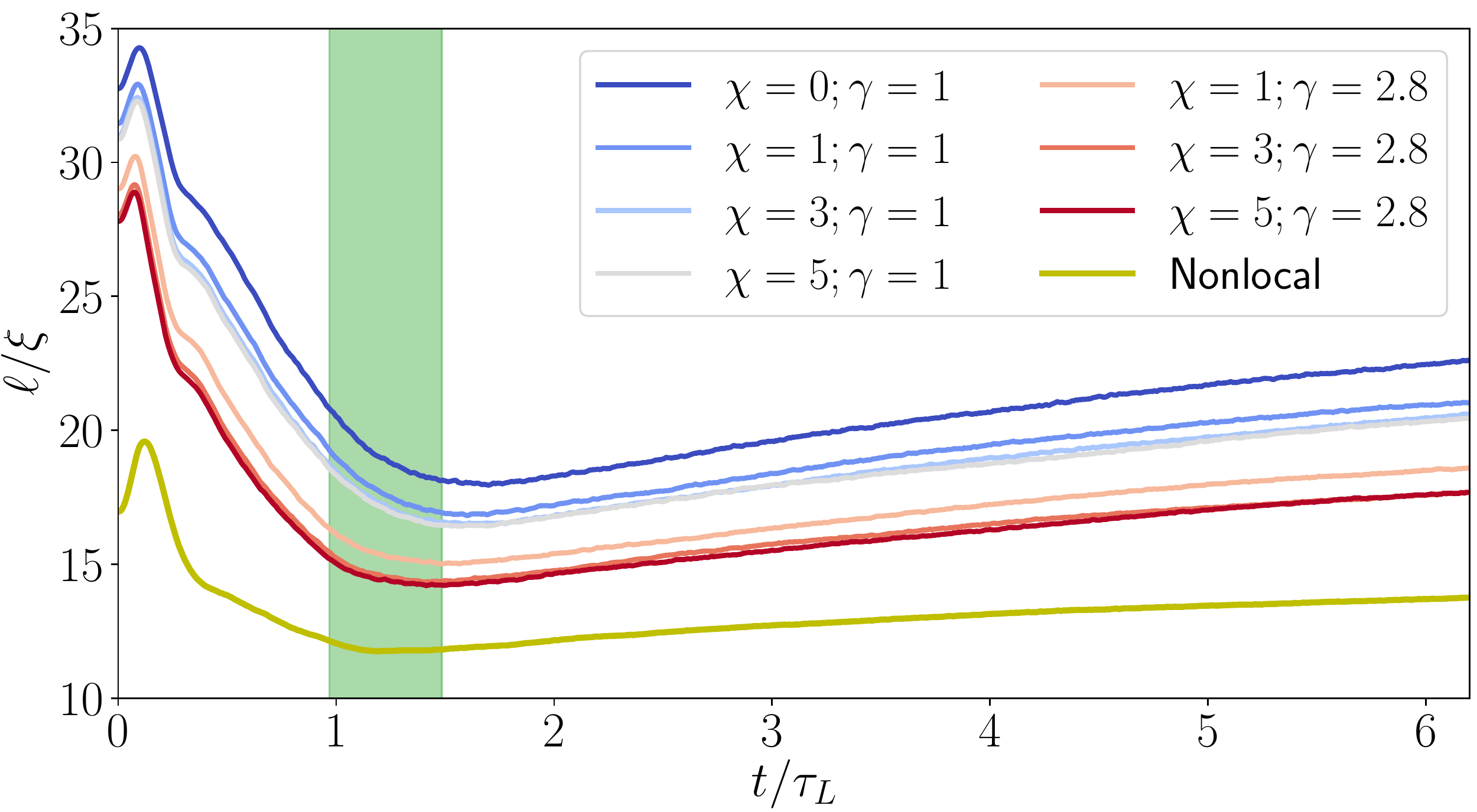}
  \caption{Time evolution of the intervortex distance of the system in
          units of the healing length.  All curves correspond to the runs A in Table \ref{tab:simulations}. The green area corresponds to the window where the dissipation achieves a maximum and where the time averages were performed.}
  \label{fig:intervortex}
\end{figure}

As a turbulent flow evolves, the total vortex length $L_{\rm v}$ varies in time in a competition between the vortex line stretching and the reconnection process. This quantity can be obtained from the incompressible momentum density of the flow $J^I(k)$ and of a two-dimensional point-vortex 
$
J_{\mathrm{vort}}^{2D}(k)$ as $ L_{\rm v} =2\pi \sum\limits_{k<k_{\mathrm{max}}}k^2 J^I(k)/\int_0^{k_{\mathrm{max}}} k^2 J_{\mathrm{vort}}^{2D}(k)\mathrm{d}k.
$
Figure \ref{fig:intervortex} shows the time evolution of the intervortex
distance $\ell=\sqrt{V/L_{\rm v}}$. In the cases of a local and a nonlocal
interaction, the intervortex distance achieves a minimum around one $\tau_L$.
The vortex line density of the system $\mathcal{L} = \ell^{-2}$ is expected to decay in time following either $\mathcal{L}\sim t^{-1}$ \cite{villois_evolution_2016} or $\mathcal{L}\sim t^{-3/2}$ \cite{Walmsley_coexistence_2017}. However, to study the decay of the vortex line density it is necessary to study the evolution of the system for long times, far away from the turbulent regime, which is out of the scope of this work. Besides, the method implemented for the computation of the vortex length of the system is just an estimation and a more precise one should be performed for the study of the scaling.

\begin{figure}[tpb]
  \centering
  \includegraphics[width=1\linewidth]{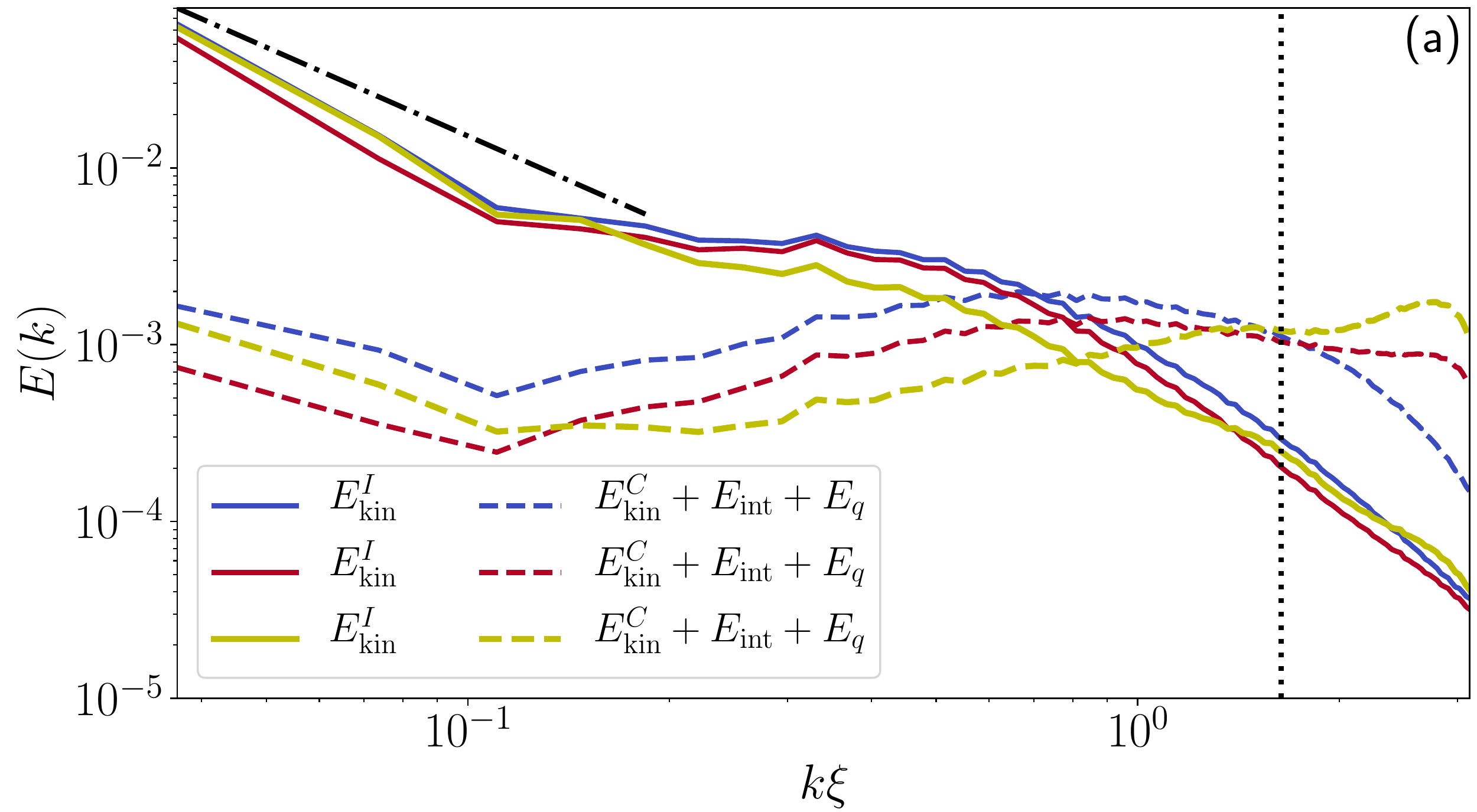}
  \includegraphics[width=1\linewidth]{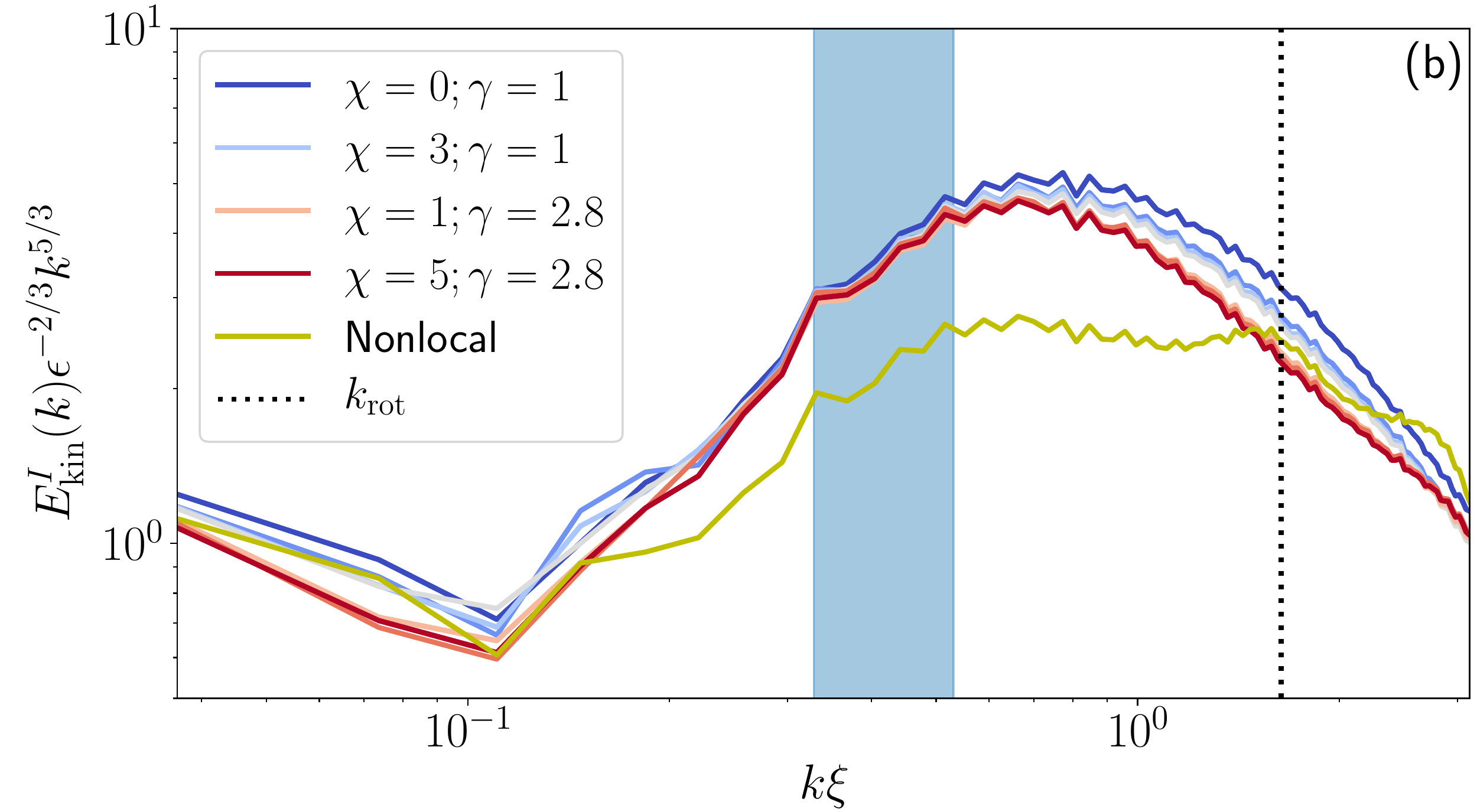}
  \caption{(a) Time averaged spectra of the incompressible kinetic energy and the sum of the internal, quantum and compressible kinetic energy for runs A1 (blue), A7 (red) and A8 (yellow) in Table \ref{tab:simulations}}. (b) Compensated incompressible energy spectra for all the set of runs A. The filled blue area indicates the intervortex wave numbers $k_{\ell} \xi$ for the different simulations.
  \label{fig:spectra}
\end{figure}

Finally, in Fig. \ref{fig:spectra} we display the energy spectra for different runs of set A.
Figure \ref{fig:spectra} (a) shows the spectra of the incompressible kinetic energy spectra and the sum of 
all the other components for different runs. Even though the range of scales is rather limited for this set of simulations, a Kolmogorov-like power law at large scales is observed in the incompressible kinetic energy. The spectra of the sum of the other energy components can be considered as the contribution of excitations that do not arise from vortices. Phenomenologically, we can consider that dynamics of the system is governed by vortices, and thus  almost incompressible, for scales down to the crossover between the two spectra plotted in Fig. \ref{fig:spectra} (a). Such crossover wave number is decreased while introducing beyond mean field terms and a nonlocal potential. Figure \ref{fig:spectra} (b), displays the incompressible energy spectra compensated by the Kolmogorov prediction $E_{\mathrm{kin}}^I \sim \epsilon^{2/3}k^{-5/3}$, where large scales collapse to values close to one. Remarkably, for the nonlocal potential run, a secondary plateau appears at smaller scales, below the intervortex distance (intervortex wave numbers for each run vary within the blue area). This range can be associated to the presence of Kelvin waves and it will be studied at higher resolutions in the next section.

\subsection{Numerical evidence of the coexistence of Kolmogorov and Kelvin wave cascades}
\label{subsec:ResultsII}

The Kelvin wave cascade discussed in Sec. \ref{subsec:cascades}, is formally derived from an incompressible model in a very simplified theoretical setting. In the context of the GP model, the Kelvin wave cascade was first observed in reference \cite{krstulovic_kelvinwave_2012} where a setting close to the theoretical prediction was used. In the case of turbulent tangles, there was first an indirect observation of the Kelvin wave cascade by making use of the spatiotemporal spectra
\cite{clarkdileoni_spatiotemporal_2015}. In that work, the Kelvin wave dispersion relation was glimpsed and a space-time filtering of the fields was performed yielding a scaling in the energy spectrum compatible to the Kelvin wave cascade. Then, by using an accurate tracking algorithm of a turbulent tangle, in reference \cite{villois_evolution_2016} the L'vov-Nazarenko prediction was clearly observed in the spectrum of large vortex rings extracted from the tangle. Later, in Refs. \cite{clarkdileoni_dual_2017,shukla_quantitative_2019}, by using high-resolution numerical simulations of the GP model, a secondary scaling range compatible with Kelvin wave cascade predictions was observed.
In this section, we focus on the scaling of the incompressible energy spectra and helicity for the case with a nonlocal potential (set of runs B) as it seems to present a much clearer scaling at scales smaller than the intervortex distance. We vary different parameters so the range of scales (system size, intervortex distance, healing length) and energy fluxes take different values.

\begin{figure}[tpb]
  \centering
  \includegraphics[width=1\linewidth]{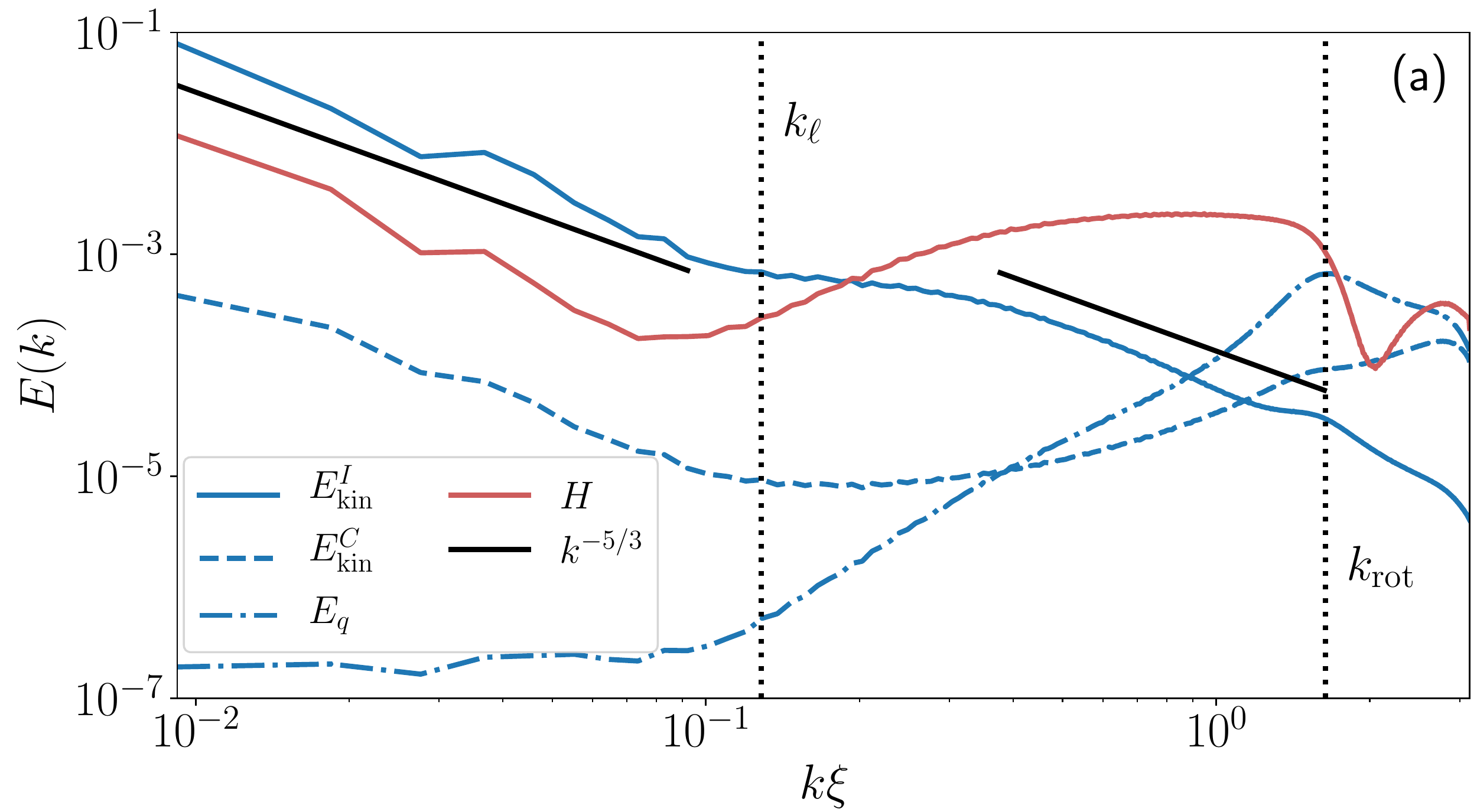}
  \includegraphics[width=1\linewidth]{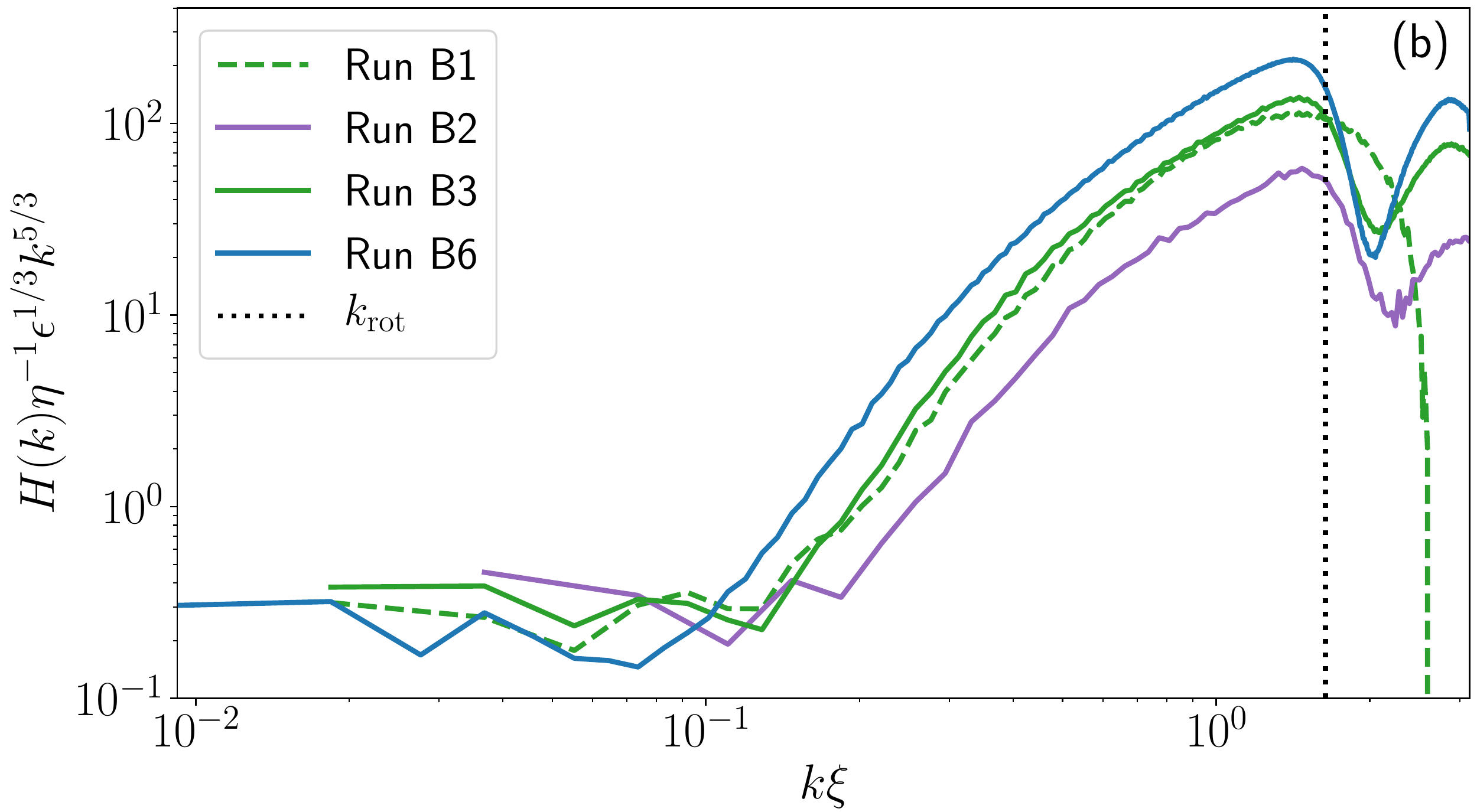}
      \caption{(a) Helicity and energy spectra of the different components for the
      simulation with $1024^3$ grid points (Run B6). Vertical dotted lines indicate the wave numbers associated with the intervortex distance $k_{\ell}$ and the roton minimum $k_{\mathrm{rot}}$.} (b) Helicity spectra compensated by Eq \eqref{eq:hel_spec} for runs B1-B3 and B6 shown in Table \ref{tab:simulations}. 
  \label{fig:spectra1024}
\end{figure}

The spectra for the different components constituting the kinetic energy and
the helicity of the simulation with $1024^3$ grid points are shown in figure
\ref{fig:spectra1024} (a).
Clear power laws for the Kolmogorov and Kelvin wave range are observed. 
A fit $k^{-\alpha}$ using the least squares method was performed for each
cascade, obtaining a scaling $\alpha=1.73$ in the range between $k\xi=0.02$ and
$k\xi=0.12$ associated with the Kolmogorov cascade and a scaling $\alpha=1.65$
in the range between $k\xi = 0.33$ and $k\xi=1.64$ associated with a Kelvin
waves cascade. These two scaling laws are separated by the intervortex
wave number $k_{\ell}$. At this scale, a bottleneck between a strong and a weak
cascade takes place and, as a consequence, a plateau in the incompressible
kinetic energy is observed. According to the warm cascade ideas \cite{lvov_bottleneck_2007},
this bottleneck should display a $k^2$ scaling associated with the thermalization
of the system. However, this behavior is not observed probably due to the fact
that the separation of scales in the Kelvin wave range is not large enough.

Concerning other energy components, the quantum energy shows a maximum at the scale associated with the roton
minimum, whereas its contribution is negligible at large scales. The helicity
spectrum also displays a Kolmogorov-like behavior at large scales, while at
scales between the intervortex distance and the roton minimum it flattens. This
flat range of the helicity spectrum appears in the range where the Kelvin wave
cascade is dominant. Whether it exists a direct relationship between the Kelvin
wave cascade and the flattening of the central line helicity spectrum, is still
unclear.
Figure \ref{fig:spectra1024} (b) displays the compensated helicity spectrum
according to \eqref{eq:hel_spec} for different runs displaying different scale
separations and with local and nonlocal potentials. The parameters of these
simulations correspond to the runs B1-B3 and B6 shown in Table
\ref{tab:simulations}. At large scales all curves collapse to a constant
$C_H\sim 1$, while at smaller scales the system with a wider scale separation
displays that the helicity contribution is more intense.

To analyze further the incompressible energy spectra, we have performed two runs varying the integral length of the initial condition so that the dissipation rate also changes (Runs B4-B5). We recall that in classical turbulence, the energy flux $\epsilon$ is fixed by the inertial range and varies as $\epsilon\sim v_{\rm rms}^3/L_0$. Our initial condition $\psi_{\rm ABC}$ keeps fixed, by construction, the value of $v_{\rm rms}$. In Table \ref{tab:turbValues} we present the values of different physical quantities relevant for a turbulent state.
\begin{table}
\centering
\begin{tabular}{c@{\hspace{1pt}}||c|c|c|c|c|c||}
Run     & $v_{\rm rms}$	& $L_0$ &$\kappa$  & $\epsilon$     &$\ell$	 	 \\
     \hline \hline
B1 & 0.395	& $L/2$ & 0.163  & 0.012   	& 0.412         \\
B2 & 0.377	& $L/2$ & 0.327  & 0.013  	& 0.494    	      \\
B3 & 0.398	& $L/2$ & 0.163  & 0.012  	& 0.255          \\
B4 & 0.406	& $L/3$ & 0.163  & 0.020  	& 0.235        	\\
B5 & 0.403	& $L/4$ & 0.163  & 0.029	& 0.227    	     \\
B6 & 0.392	& $L/2$ & 0.081  & 0.011  	& 0.139 	      \\
\hline
\end{tabular}
\caption{Values of integral scale $L_0$, the quantum of circulation $\kappa$, the energy dissipation rate $\epsilon$ and the intervortex distance, expressed in units of the box size $L=2\pi$ and the speed of sound is fixed to $c=1$.}
\label{tab:turbValues}
\end{table}
 Such quantities are expressed, as customary in classical turbulence, in units of large scale quantities. In particular, the system size is $L=2\pi$ and the speed of sound is $c=1$. With such definitions, large scale quantities remain almost constant when increasing the scale separation between the box size and the smallest scale of the system, but the quantum of circulation takes smaller values.

\begin{figure}[tpb]
  \centering
  \includegraphics[width=1\linewidth]{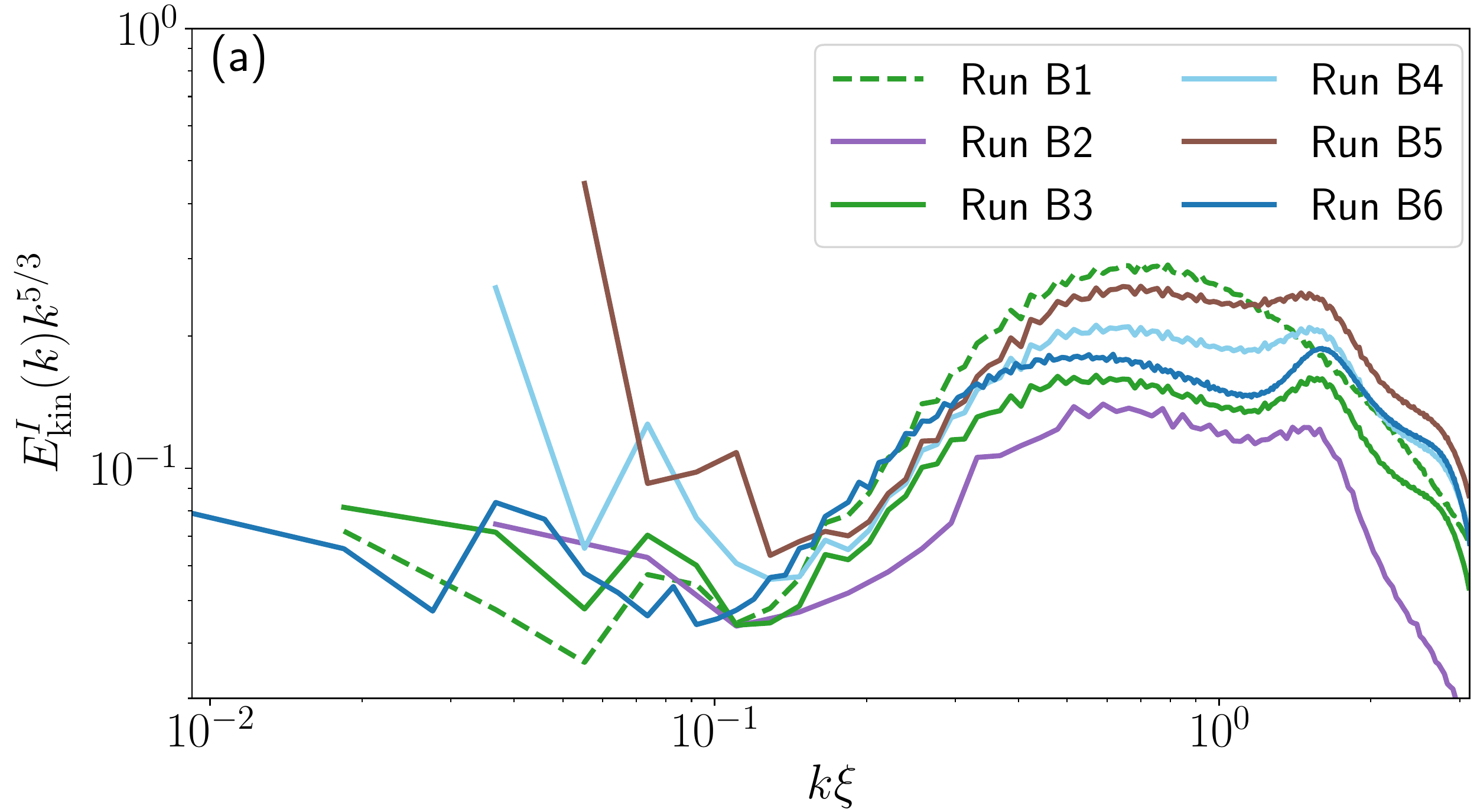}
  \includegraphics[width=1\linewidth]{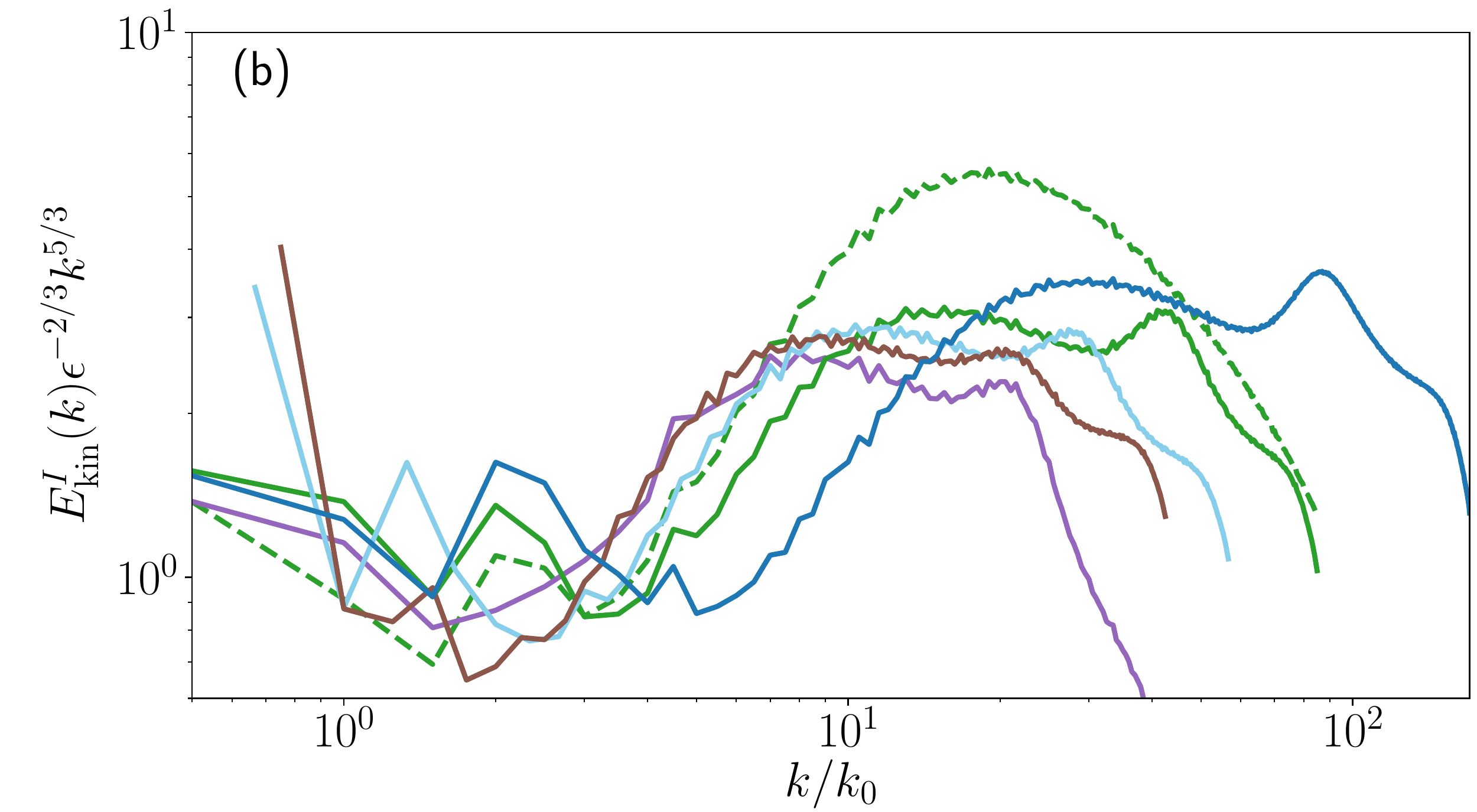}
  \includegraphics[width=1\linewidth]{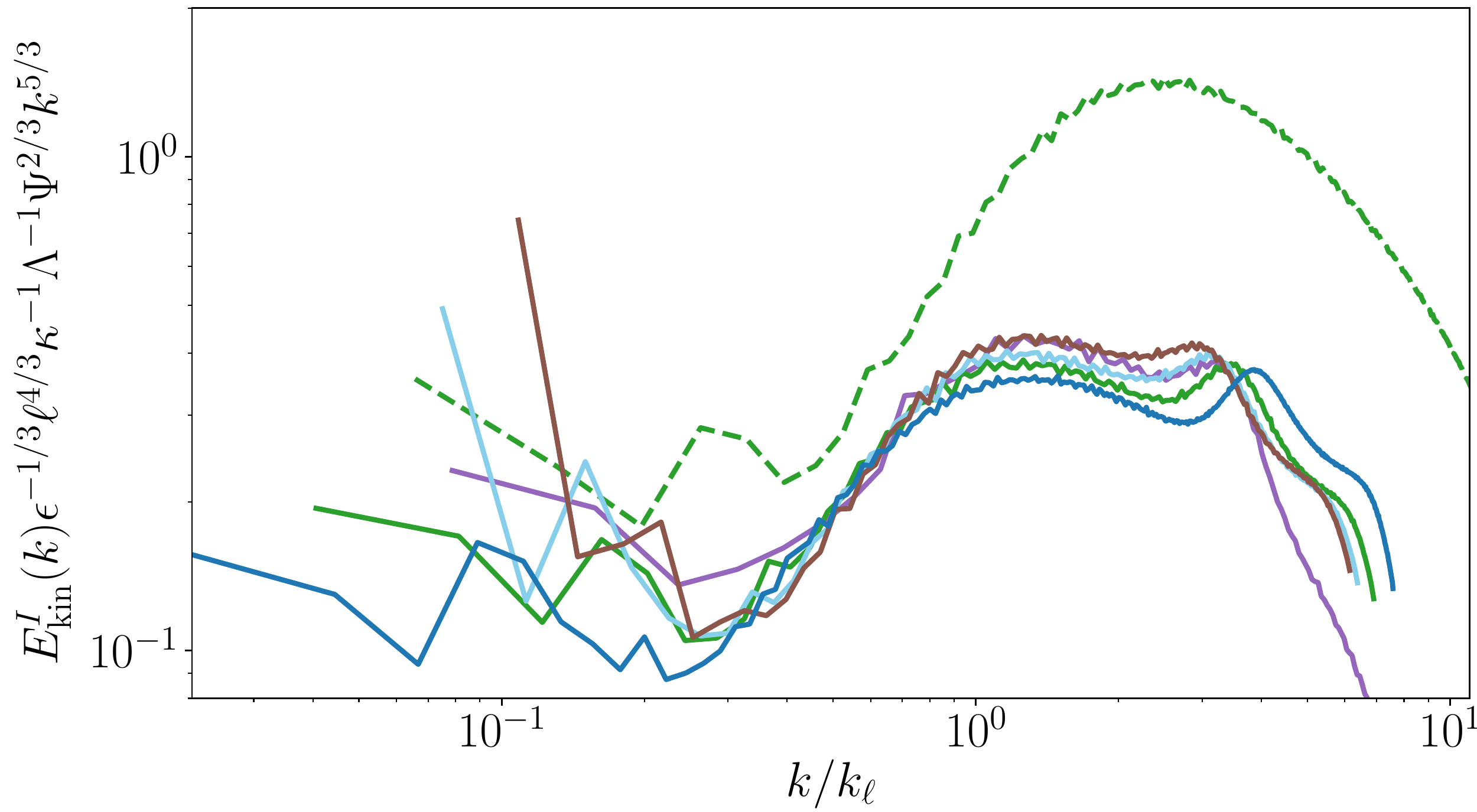}
  \caption{Compensated incompressible kinetic energy spectra by (a) $k^{-5/3}$ scaling, (b) Kolmogorov scaling \eqref{eq:K41} and (c) L'vov-Nazarenko scaling for Kelvin waves \eqref{eq:KW}.}
  \label{fig:compensatedK41}
\end{figure}

Figure \ref{fig:compensatedK41} (a) shows the incompressible energy spectra
simply compensated by $k^{-5/3}$.  Two plateaux are clearly observed, one
corresponding to the large scales Kolmogorov scaling and the other to the small
scales Kelvin wave cascade. It can be seen that the values of these plateaux
differ by a factor of order 3. By following the procedures introduced in reference
\cite{lvov_bottleneck_2007} but using Eq.~\eqref{eq:KW} for the scaling of
Kelvin waves, the ratio between these two plateaux is expected to scale with
$\Lambda$ as $E_{KW}/E_{K41} \sim \Lambda^{22/15}$. In our simulations, the
ratio between the ratio of energies and $\Lambda^{22/15}$ is in average close
to $0.1$. It is important to have in mind that this prediction is expected to
take place in the limit of $\Lambda \gg 1$, that is, a huge scales separation
between $\ell$ and $\xi$. In our simulations this quantity takes an average
value of $\Lambda \approx 2.7$
.  

The energy spectra shown in Fig. \ref{fig:compensatedK41} (b) have been compensated by the Kolmogorov law \eqref{eq:K41} and displayed as a function of $k/k_0$, with $k_0=2\pi/L_0$ in order to emphasize the Kolmogorov regime. Once properly normalized, all runs present a plateau at large scales that collapse to values that fluctuate around a Kolmogorov constant $C_K\sim1$, in agreement with previous simulations of the GP model \cite{clarkdileoni_dual_2017,shukla_quantitative_2019}. In order to emphasize the Kelvin wave cascade, we make use of the L'vov \& Nazarenko wave turbulence prediction \eqref{eq:KW}. Figure \ref{fig:compensatedK41} (c) displays the incompressible energy spectra compensated by this theoretical prediction as a function of $k/k_\ell$, with $k_\ell=2\pi/\ell$. The collapse of the Kelvin wave cascade is remarkable. All runs having a nonlocal potential display a plateau around a value $C_{\mathrm{LN}}^{3/5} \approx 0.36$, which recovers a constant of $C_{\mathrm{LN}} \approx 0.18$. Such value is relatively close to the predicted one $C_{\mathrm{LN}}=0.304$, in particular considering all the phenomenological assumptions made in Sec. \ref{subsec:cascades} to adapt the theoretical prediction \eqref{eq:KW_badunits} to the case of a turbulent tangle in Eq. \ref{eq:KW}. It is also important to remark that Eq. \ref{eq:KW_badunits} is obtained from a local Biot-Savart model, while the dynamics studied in this work corresponds to the gGP model, with a nonlocal interaction potential and beyond mean field corrections.
The GP run (with local interaction potential) displays a good Kolmogorov scaling at large scales. However, it does not clearly exhibit a Kelvin wave cascade range at the highest resolution used in this work for this model  ($512^3$ grid points). Note that previous works reporting a secondary $k^{-5/3}$ range in local GP model have used resolutions of $2048^3$ \cite{clarkdileoni_dual_2017} and $4096^3$ \cite{shukla_quantitative_2019} collocation points. The incompressible kinetic energy spectrum, compensated by the Kozik \& Svistunov prediction \cite{kozik_theory_2009} is displayed in Appendix \ref{sec:KSspectrum}.

\section{Conclusions\label{sec:conclusions}}

We studied the properties of the freely decaying quantum turbulence of the
generalized Gross-Pitaevskii (gGP) model \eqref{eq:gGPE}, that includes beyond
mean field corrections and considers a nonlocal interaction potential between
bosons. This model pretends to give a better description of superfluid helium as it
reproduces a roton minimum in the excitation spectrum. 

The visualization of the flow with a nonlocal potential allowed us to observe
the formation of helical structures around the vortices produced by density
fluctuations, exhibiting the intrinsic property of maximal helicity of an ABC
flow. These structures were not observed at initial times in a flow with no
helicity like a Taylor-Green flow, but they develop as the system evolves (data
not shown). However, it was seen that the behavior of the helicity is
independent of the interaction potential. At large scales the helicity develops
a spectrum that satisfies prediction \eqref{eq:hel_spec}, while at scales
between the intervortex distance and the healing length a plateau is observed.
This range is usually associated with the Kelvin wave cascade regime, but it is
still not clear whether the formation of this plateau is associated with Kelvin
waves or not. 

By studying numerically the freely decaying quantum turbulence of an ABC flow,
we observed that the statistical behavior of the system 
does not depend much on the parameters of the beyond mean field
correction in the presence a local interaction potential between bosons. 
This is manifest in the evolution of quantities such as the energy, the helicity and the intervortex distance of the system.
The introduction of a nonlocal potential does not modify significantly the
behavior of the system at large scales, exhibiting a Kolmogorov-like scaling
law for the incompressible kinetic energy. However, the situation changes at
smaller scales when a nonlocal potential is implemented, between the
intervortex distance $\ell$ and the healing length $\xi$, range associated with
the Kelvin waves cascade. Here, the nonlocal potential and roton minimum enhance a second scaling of the incompressible energy
spectrum. This is observed even at a moderate resolution of $256^3$ grid points,
while in the case of a local GP model an energy spectrum compatible with
$k^{-5/3}$ scaling law begins to be recognizable from resolutions of $2048^3$
collocation points \cite{clarkdileoni_dual_2017}, and even in this case the
range of scales where it takes place is less than a decade. This stronger manifestation of the Kelvin waves cascade may be very
useful for numerical and theoretical studies of wave turbulence.  This clear
difference with the local GP model may be used to compare if effectively this
model better describes the dynamics of superfluid helium.  However,
experimental observation at scales smaller than the intervortex distance still
remains a challenge.

We also studied how is the scaling of the Kelvin wave spectrum with the energy
flux $\epsilon$ and the intervortex distance by varying the integral scale of
the initial flow and its healing length. We observed that the different spectra
tend to collapse to a constant according to L'vov \& Nazarenko spectrum for
Kelvin waves \eqref{eq:KW}. The observed value of the constant is
$C_{\mathrm{LN}} \approx 0.18$ which is close to the predicted one
$C_{\mathrm{LN}} \approx 0.304$. This shows how robust this prediction is in
the presence of a nonlocal interaction potential. This is surprising given that
the theory is constructed using a local Biot-Savart model considering a single
vortex line while here it is extended to a vortex tangle in the framework 
of the gGP model with a nonlocal interaction potential and including
several phenomenological assumptions. The Kozik \& Svistunov spectrum for
Kelvin waves was also studied for these set of simulations, however, by
compensating the energy spectra by this theory no clear plateau is observed (see Appendix
\ref{sec:KSspectrum}). Furthermore, in the range of the Kelvin wave cascade the
Kozik-Svistunov cascade would take values of $C_{\mathrm{KS}} \approx 0.06$ which
is not of order one, so it might imply that the energy spectrum of this system
is not described by this theory. 

The overall results of this work show that, both GP and gGP models describe a
similar behavior at large scales, including a Kolmogorov-like spectrum of the
incompressible kinetic energy and the scaling law observed for the helicity.
However, at small scales the gGP includes the roton minimum in the excitation
spectrum and a Kelvin waves cascade is enhanced, showing a clear discrepancy
with the local GP model.  This means that the model used to describe the
interaction between bosons only affects the dynamics of the system at 
scales smaller than the intervortex distance.  

%
%
%

\appendix
\section{Kozik-Svistunov Kelvin spectrum \label{sec:KSspectrum}}

\begin{figure}[h!]
  \centering
  \includegraphics[width=1\linewidth]{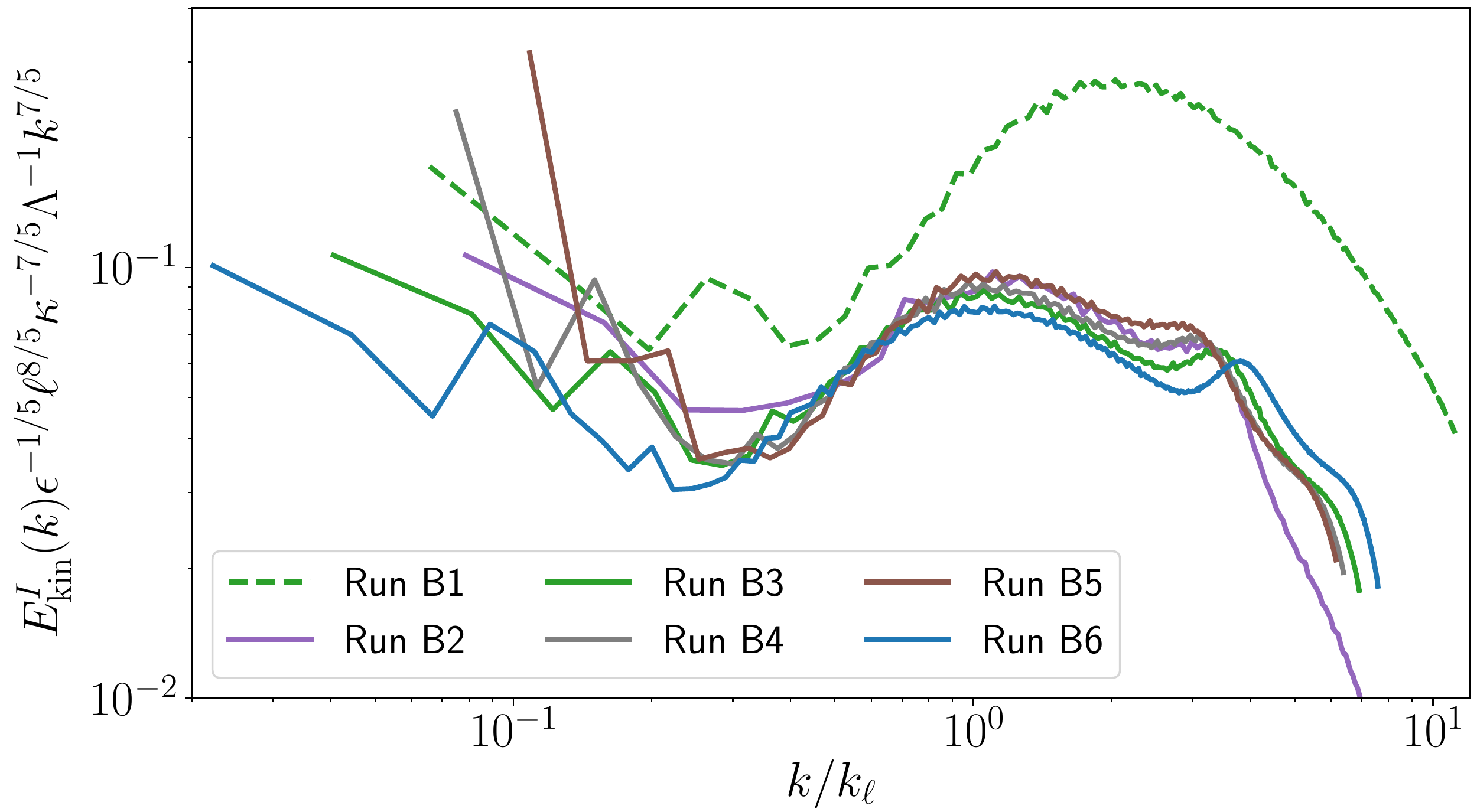}
  \caption{Compensated incompressible kinetic energy spectra by the Kozik \& Svistunov prediction for Kelvin waves.}
  \label{fig:compensatedKS}
\end{figure}

The original Kozik \& Svistunov prediction for the Kelvin wave cascade  \cite{kozik_theory_2009} was done with same geometrical considerations of L'vov \& Nazarenko and also expressed in units of $Length^5/Time^2$. Applying the same considerations of Sec. \ref{subsec:cascades} to adapt this prediction to a turbulent three-dimensional flow leads to the following Kelvin wave energy spectrum
\begin{equation}
        E_{\mathrm{KW}}^{\rm KS}(k) = C_{\mathrm{KS}}
        \frac{\kappa^{7/5}\Lambda\epsilon^{1/5}\ell^{-8/5}}{k^{7/5}}.
        \label{eq:KWKS}
\end{equation}
where the constant $C_{\mathrm{KS}}$ could be in principle determined by the theory if some integrals in the associated kinetic equation are convergent, but its value is still unknown. Figure \ref{fig:compensatedKS} displays the incompressible kinetic energy spectrum compensated by prediction \eqref{eq:KWKS}.
All the curves tend to collapse in the range associated with Kelvin waves,
showing a proper scaling with the energy flux $\epsilon$, the intervortex
distance $\ell$ and the quantum of circulation $\kappa$.  However, although the
Kelvin wave range is limited, a plateau is not clearly observed if the spectra
are compensated by \ref{eq:KWKS} and even though a constant cannot be well
defined, the energy spectra collapse to a constant of $C_{\mathrm{KS}} \approx
0.06$, which is not of order one.

\begin{acknowledgments}
N.P.M and G.K were supported by Agence Nationale de la Recherche through the project GIANTE ANR-18-CE30-0020-01. Computations were carried out on the M\'esocentre SIGAMM hosted at the Observatoire de la C\^ote d’Azur and the French HPC Cluster OCCIGEN through the GENCI allocation A0042A10385. GK is also supported by the EU Horizon 2020 Marie Curie project HALT and the Simons Foundation Collaboration grant Wave Turbulence (Award ID 651471).
\end{acknowledgments}

\bibliographystyle{apsrev4-2}

\bibliography{bibliography}

\end{document}